\documentclass[a4paper,fleqn,usenatbib]{mnras}
\usepackage[T1]{fontenc}
\usepackage{ae,aecompl}
\usepackage{graphicx}
\usepackage{amsmath}
\usepackage{amssymb}
\usepackage{txfonts}
\usepackage{float}
\usepackage{adjustbox}
\usepackage{siunitx}
\usepackage{xcolor}

\usepackage{tensor}

\graphicspath{{figures/}}

\newcommand{\ltsima}{$\; \buildrel < \over \sim \;$}
\newcommand{\lsim}{\lower.5ex\hbox{\ltsima}}
\newcommand{\gtsima}{$\; \buildrel > \over \sim \;$}
\newcommand{\gsim}{\lower.5ex\hbox{\gtsima}}
\newcommand{\bra}{\langle}
\newcommand{\ket}{\rangle}

\newcommand{\dd}{\mathrm{d}}

\newcommand{\likeli}{\mathcal{L}}

\title[physical MCMC convergence criteria]
{Partition function approach to non-Gaussian likelihoods: physically motivated convergence criteria for Markov-chains}
\author[L. R{\"o}ver, H. v. Campe, M.Ph. Herzog, R.M. Kuntz, B.M. Sch{\"a}fer]
{Lennart R{\"o}ver$^{1,2}$\thanks{e-mail: l.roever@stud.uni-heidelberg.de}, Heinrich von Campe$^2$,  Maximilian Philipp Herzog$^2$,\newauthor Rebecca Maria Kuntz$^2$, Bj{\"o}rn Malte Sch{\"a}fer$^2$\thanks{e-mail: bjoern.malte.schaefer@uni-heidelberg.de}\\
$^1$ Institut f{\"u}r Theoretische Physik, Universit{\"a}t Heidelberg, Philosophenweg 16, 69120 Heidelberg, Germany\\
$^2$Zentrum f{\"u}r Astronomie der Universit{\"a}t Heidelberg, Astronomisches Rechen-Institut, Philosophenweg 12, 69120 Heidelberg, Germany
}

\begin{document}
\pagerange{\pageref{firstpage}--\pageref{lastpage}}
\pubyear{2023}
\maketitle
\label{firstpage}

\begin{abstract}
Non-Gaussian distributions in cosmology are commonly evaluated with Monte Carlo Markov-chain methods, as the Fisher-matrix formalism is restricted to the Gaussian case. The Metropolis-Hastings algorithm will provide samples from the posterior distribution after a burn-in period, and the corresponding convergence is usually quantified with the Gelman-Rubin criterion. In this paper, we investigate the convergence of the Metropolis-Hastings algorithm by drawing analogies to statistical Hamiltonian systems in thermal equilibrium for which a canonical partition sum exists. Specifically, we quantify virialisation, equipartition and thermalisation of Hamiltonian Monte Carlo Markov-chains for a toy-model and for the likelihood evaluation for a simple dark energy model constructed from supernova data. We follow the convergence of these criteria to the values expected in thermal equilibrium, in comparison to the Gelman-Rubin criterion. We find that there is a much larger class of physically motivated convergence criteria with clearly defined target values indicating convergence. As a numerical tool, we employ physics-informed neural networks for speeding up the sampling process.
\end{abstract}


\onecolumn

\section{introduction}
The basis for inference of model parameters $\theta$ from data $y$ is Bayes' theorem \citep[for applications to cosmology, see][among many others]{trotta_bayesian_2017, trotta_bayes_2008}: It links likelihood $\mathcal{L}(y|\theta)$ and prior $\pi(\theta)$ to the posterior distribution $p(\theta|y)$,
\begin{equation}
p(\theta|y) = \frac{\mathcal{L}(y|\theta)\pi(\theta)}{p(y)}
\quad\text{with the Bayesian evidence}\quad
p(y) = \int\dd^n\theta\:\mathcal{L}(y|\theta)\pi(\theta)
\end{equation}
as the normalisation of the posterior distribution. It is well-known that the evidence can be extended to a canonical partition sum $Z[\beta,J_\alpha]$
\begin{equation}
Z[\beta,J_\alpha] = 
\int\dd^n\theta\:\left(\mathcal{L}(y|\theta)\pi(\theta)\right)^\beta\exp(\beta J_\alpha\theta^\alpha)
\end{equation}
with an inverse temperature $\beta$. Associated to the canonical partition is the Helmholtz energy $F[\beta,J_\alpha]$
\begin{equation}
F[\beta,J_\alpha] = \frac{1}{\beta}\ln Z[\beta,J_\alpha]
\end{equation}
from which cumulants of the posterior distribution $p(\theta|y)$ follow by differentiation with respect to $J_\alpha$, evaluated at $J_\alpha=0$ and $\beta=1$. 

For the particular case of Gaussian likelihoods the Fisher-matrix formalism \citep{tegmark_karhunen-loeve_1997} provides a very powerful analytical tool with many applications in cosmology \citep{coe_fisher_2009, bassett_fisher_2011, wolz_validity_2012, elsner_fast_2012, raveri_cosmicfish_2016}, with extensions to weak non-Gaussianities \citep{sellentin_breaking_2014, sellentin_fast_2015,schafer_describing_2016}. Properly non-Gaussian likelihoods and posterior distributions require the use of Monte Carlo Markov-chains. Introduced by \citet{lewis_cosmological_2002} in cosmology, Markov-chains generate samples $\theta^{(i)}$ from the posterior distribution $p(\theta|y)$ by virtue of the Metropolis-Hastings algorithm \citep{metropolis_monte-carlo:_1985, metropolis_equation_1953}. This algorithm simulates a thermal random walk at temperature $1/\beta$ on a potential $\Phi(\theta) = -\ln(\mathcal{L}(y|\theta)\pi(\theta))$. For Gaussian error processes, this effective potential $\Phi(\theta) = \chi^2(y|\theta)/2 + \phi(\theta)$ is composed of $\chi^2(y|\theta)$, of the fit, since $\mathcal{L}\propto\exp(-\chi^2(y|\theta)/2)$, as well as the logarithmic prior $\phi(\theta)$.

But the initial samples of the Metropolis-Hastings algorithms do not follow the posterior distribution $p(\theta|y)$. It is only after a burn-in period, whose length depends on the particular shape of the likelihood as well as the starting point of the Markov-chain, that the sampling is fair and representative of the underlying distribution \citep{10.1214/ss/1015346320, 10.1214/aos/1176325750}. The convergence of the sampling can be quantified with the Gelman-Rubin criterion \citep{gelman_rubin}, which ensures that the covariance of samples from a single Markov-chain and the covariance between distinct Markov-chains at the same instant are equal, and that the samples drawn in the Markov-process allow computing integrals weighted with the posterior distribution. 

While this is certainly sensible, we try to characterise the burn-in phase and the equilibration of the Markov-chain with criteria that are motivated by statistical mechanics, driven by the physical picture of the Markov-chain performing a thermal random walk inside a potential. The equilibrium values of these criteria are computable from canonical partition sums, such that the target values of these criteria to be reached in burn-in are known.

Even though many of our results would apply to conventional Metropolis-Hastings algorithms, we focus on Hamilton Monte Carlo-samplers \citep{DUANE1987216, Neal2011MCMCUH}, not only because they can be more efficient in cases of degeneracies but also because they have a notion of total energy, with a potential and a kinetic contribution. We would argue that the Markov-chain first reaches a state of virialisation, where kinetic and potential energy are in a certain fixed ratio to each other that depends on the shape of the potential. Then, the Markov-chain establishes equipartition, where the coordinates as degrees of freedom become independent of each other and the energies associated with every degree of freedom assume the same value proportional to temperature.

Furthermore, the canonical partition $Z[\beta,J_\alpha]$ assumes the existence of a heat bath at temperature $1/\beta$. As the Metropolis-Hastings algorithm reaches thermodynamic equilibrium, the exchange of thermal energy with the environment subsides up to thermal fluctuations. All three conditions, virialisation, equipartition and thermal equilibrium should be viable, physically motivated conditions to characterise the convergence of Markov-chains. The central motivation of our paper is therefore to understand the equilibration, or burn-in process of Markov-chains by drawing an analogy to statistical physics in thermal equilibrium, and comparing quantities in statistical mechanics with a well-defined equilibrium value with their analogous quantities in statistical inference. It is worth pointing out that the Gelman-Rubin criterion is formulated as a statistical test for the equality of two variances (one defined as an ensemble average and the other as an average for subsequent samples of a single chain), and that the virialisation, equipartition and thermalisation conditions can likewise be formulated as statistical tests, in this case for equal mean: This is possible, because statistical mechanics in thermodynamic equilibrium makes specific predictions about the numerical value of these quantities, so the target values that an equilibrated chain should reach are well-defined.

Throughout the paper, we work with the summation convention and denote parameter tuples $\theta^\mu$ and data tuples $y^i$ as vectors with contravariant indices; Greek letters and indices are reserved for quantities in parameter space and Latin letters and indices for objects in data space. With these conventions, the data covariance $C^{ij} = \bra y^i y^j\ket-\bra y^i\ket\bra y^j\ket$ is a contravariant tensor with $C_{ij}$ as its covariant inverse, $C^{ij}C_{jk} = \delta^i_k$. Applying the identical definitions, the Fisher-matrix $F_{\alpha\beta}$ is a covariant tensor defining quadratic forms through $\chi^2 = F_{\alpha\beta}\theta^\alpha\theta^\beta$. The inverse Fisher matrix $F^{\alpha\beta}$, determined by $F_{\alpha\beta}F^{\beta\gamma} = \delta_\alpha^\gamma$ would then correspond to the parameter covariance of a multivariate Gaussian distribution.

After introducing the partition function formalism for Monte Carlo Markov-chains in Sect.~\ref{sect:mcmc} we discuss the properties of virialisation in Sect.~\ref{sect:virialisation}, the stationarity of thermal ensembles in Sect.~\ref{sect:stationarity} before we come to conditions characterising Sect.~\ref{sect:thermalisation}. We carry out a number of numerical experiments with likelihoods that are known to be difficult to sample as they are far from Gaussianity before investigating convergence of MCMC-sampling in the supernova likelihood as a physical example (Sect.~\ref{sect:supernovae}), with the parameter space formed by the dark matter density $\Omega_m$ and the dark energy equation of state $\omega$, for flat Friedmann-cosmologies. We summarise our main findings in Sect.~\ref{sect_summary}.

\section{Partition functions and Hamilton Monte Carlo sampling}\label{sect:mcmc}
\label{sect:partitions_MCMC}
Similar to \cite{Rover:2022mao} a partition function can be assigned to a particular posterior distribution through the associated evidence by introducing a source $J_\alpha$ and an inverse temperature $\beta$
\begin{align}
	Z[\beta,J_{\alpha}] = 
	\int\dd^n\theta\: \exp\left(-\beta\left[\frac{1}{2}\chi^2(y|\theta) + \phi(\theta)\right]\right) \exp\left(\beta J_{\alpha}\theta^{\alpha}\right),
	\label{eq:Z_no_kin}
\quad\text{with}\quad
\mathcal{L}(y|\theta)\propto\exp\left(-\frac{\chi^2(y|\theta)}{2}\right)
\quad\text{and}\quad
\pi(\theta)\propto\exp(-\phi(\theta)),
\end{align}
where the logarithmic likelihood and prior play the role of a potential, specifically $\Phi(\theta) = \chi^2(y|\theta)/2 + \phi(\theta)$. The partition function then describes a sum over the potential energies associated to all possible microstates, which in this case correspond to parameter tuples $\theta^\mu$.

Extending the partition function with a kinetic term allows to introduce a kinetic energy $T(p)$ to each particle. For a given position $\theta^\mu$ and conjugate momentum $p_\mu$ the microscopic energy can be described using the Hamiltonian function $\mathcal{H}(p,\theta)$ \citep{liu_book, betancourt2018conceptual}
\begin{align}
	\mathcal{H}(p,\theta) = T(p) +  \Phi(\theta).
\end{align}

This Hamiltonian function can then be used to define a new partition function of the form
\begin{align}
	Z[\beta,J_{\alpha}, K^\alpha] = 
	\int\dd^n p \int\dd^n\theta\: \exp\left(-\beta \mathcal{H}(p,\theta) \right) \exp\left(\beta J_{\alpha}\theta^{\alpha}\right)\exp\left(\beta K^{\alpha}p_{\alpha}\right).
	\label{eq:Z_with_kin}
\end{align}
with analogous sources $K^\alpha$ for the canonical momenta $p_\alpha$. As the energies $\mathcal{H}(p,\theta)$ are constructed additively from the kinetic term $T(p)$ and the potential term $\Phi(\theta) = \chi^2/2+\phi$, the partition function separates
\begin{equation}
Z[\beta,J_\alpha,K^\alpha] = 
\int\dd^n\theta\: \exp\left(-\beta \Phi(\theta) \right) \exp\left(\beta J_{\alpha}\theta^{\alpha}\right) \times 
\int\dd^n p\: \exp\left(-\beta T(p) \right) \exp\left(\beta K_{\alpha}p^{\alpha}\right) = 
Z_{\Phi}[\beta,J_\alpha] \times Z_{T}[\beta,K_\alpha]
\end{equation}
and its logarithm
\begin{equation}
\ln Z[\beta,J_\alpha,K^\alpha] = \ln Z_{\Phi}[\beta,J_\alpha] + \ln Z_{T}[\beta,K^\alpha]
\end{equation}
can be used as a generating function for both the cumulants of the posterior distribution $p(\theta|y)$ in configuration space with the prescription
\begin{equation}
	\kappa^{\mu_1,\hdots, \mu_n}_\Phi = 
	\frac{\partial^n}{\partial J_{\mu_1}\hdots\partial J_{\mu_n}} \left.\left(\frac{1}{\beta} \ln Z[\beta, J_\alpha, K^\alpha]\right)\right|_{J=0, K=0, \beta= 1} =
	\frac{\partial^n}{\partial J_{\mu_1}\hdots\partial J_{\mu_n}} \left.\left(\frac{1}{\beta} \ln Z_\Phi[\beta, J_\alpha]\right)\right|_{J=0, \beta= 1}
\end{equation}
as well as the cumulants of the posterior distribution in momentum space $p(p|y)$ using the form
\begin{equation}
	\kappa_T^{\mu_1,\hdots, \mu_n} = 
	\frac{\partial^n}{\partial K^{\mu_1}\hdots\partial K^{\mu_n}} \left.\left(\frac{1}{\beta} \ln Z[\beta, J_\alpha, K^\alpha]\right)\right|_{J=0, K=0, \beta= 1}= 
	\frac{\partial^n}{\partial K^{\mu_1}\hdots\partial K^{\mu_n}} \left.\left(\frac{1}{\beta} \ln Z_T[\beta, K^\alpha]\right)\right|_{K=0, \beta= 1}
\end{equation}
The cumulants of the distributions in configuration space and momentum space are completely independent due to the factorisation of the partition sum, implying that, despite the fact that Metropolis-Hastings-like algorithms would sample from a joint posterior $p(\theta,p|y)$, marginalisation over the distribution in momentum space would trivially yield the posterior distribution of the parameters as the distribution factorises:
\begin{equation}
\int\dd^np\:p(\theta,p|y) = \int\dd^np\:p(\theta|y)p(p|y) = p(\theta|y)\int\dd^np\:p(p,\theta) = p(\theta|y)
\end{equation}
Despite the fact that the complexity of the system is increased from being $n$-dimensional to being $2n$-dimensional, the kinetic term is able to make sampling more efficient in cases with strong anisotropies of the potential term $\Phi(\theta)$. Strong statistical degeneracies of the likelihood might even suggest the choice of a prior in momentum space: Such a prior $\pi(p)$ would likewise not change the posterior distribution $p(\theta|y)$ but could be set up to make sampling more efficient by covering the degeneracies in parameter space more efficiently with samples compared to random, diffusive motion.

The physical interpretation of the canonical partition $Z[\beta,J_\alpha,K^\alpha] = Z_T[\beta,K^\alpha]\times Z_\Phi[\beta,J_\alpha]$ would be a classical (i.e. non-relativistic), ideal gas in thermal equilibrium inside a potential $\Phi$. In the following the kinetic term is always chosen as $T(p) = p^2/(2m)$, rendering the partition sum Gaussian, and enabling the usage of mathematically convenient Gaussian integrals. The parameter $m$ corresponds to the particle mass and would in this context be a numerical parameter that can be chosen out of practicality: Here, we will set $m=1$. It is well known \citep{betancourt2018conceptual} that generalisation to a symmetric and positive definite quadratic form can yield numerical advantages: For illustration, a parabolic likelihood $\chi^2 = F_{\alpha\beta}\theta^\alpha\theta^\beta$ with a Fisher-matrix $F_{\alpha\beta}$ would lead in this case to a canonical partition
\begin{equation}
Z[\beta,J_\alpha,K^\alpha] = 
\int\dd^n\theta\int\dd^np\:\exp\left(-\frac{\beta}{2}M^{\alpha\beta}p_\alpha p_\beta\right)\exp\left(-\frac{\beta}{2}F_{\alpha\beta}\theta^\alpha\theta^\beta\right)\exp\left(\beta J_\alpha\theta^\alpha\right)\exp\left(\beta K^\alpha p_\beta\right).
\label{eqn_anisotropic_hmc}
\end{equation}
A choice of $M^{\alpha\beta}$ proportional to the inverse Fisher matrix $F^{\alpha\beta}$ could be convenient, as the inverse Fisher-matrix encodes the covariance of the distribution. Then, $M^{\alpha\beta}$ is not only   symmetric but also positive definite, and would assign a low inertia to motion in the directions in which the distribution is broad. This anisotropic motion would make sampling more efficient, similarly to anisotropic proposal distribution as in affine-invariant sampling \citep{foreman-mackey_emcee:_2013, hou_affine-invariant_2012}. The seemingly capricious choice of covariant indices for $J_\alpha$ of contravariant indices for $K^\alpha$ is determined by writing the coordinate tuple $\theta^\alpha$ as a vector, implying that the velocities $\dot{\theta}^\alpha$ should be vectors. Then, the conjugate momenta $p_\alpha$ have to be linear forms, as they are given by $p_\alpha = \partial\mathcal{L}/\partial\dot{\theta}^\alpha$ with the Lagrange function $\mathcal{L}$. This in turn determines that $J_\alpha$ are linear forms and $K^\alpha$ vectors, as they are used to form the scalars $J_\alpha\theta^\alpha$ and $K^\alpha p_\alpha$, respectively.

Hamilton Monte Carlo algorithms as extension to the Metropolis-Hastings algorithms \citep{metropolis_equation_1953, metropolis_monte-carlo:_1985, 2011} have been proposed by \citet{DUANE1987216, betancourt2018conceptual}, and their working principles are thoroughly reviewed in \citet{Neal2011MCMCUH, jasche_fast_2010}: HMC improves the performance in particular in cases with strong degeneracies by alternating between jumps in parameter space analogous to Metropolis-Hastings sampling and deterministic motion where energy is conserved as a consequence of the Hamilton equations of motion. Trajectories with conserved energy are collections of microstates of equal posterior probability.

\section{virialisation}\label{sect:virialisation}
Bounded motion inside a potential exhibits the relation 
\begin{equation}
\left\langle\theta^\mu\frac{\partial\mathcal{H}}{\partial\theta^\mu}\right\rangle = 
\left\langle p_\mu\frac{\partial\mathcal{H}}{\partial p_\mu}\right\rangle
\end{equation}
which translates to a relation $2\bra T\ket = k\bra\Phi\ket$ between the average kinetic and potential energies for Hamiltonian functions that are homogeneous of order 2 in $p$ and of order $k$ in $\theta$. For ergodic systems it would not matter whether the averages are taken over time or over a statistical ensemble. As a Markov-chain starts exploring the potential $\Phi$ one would not expect that the virial relation holds straight away. Rather, only if samples over a few dynamical timescales of the systems are collected, the virial relation can be expected to hold.

For equilibrated Markov-chains which are a proper realisation of the canonical partition $Z[\beta,J_\alpha,K^\alpha]$ one can compute the expectation values in the virial theorem to be
\begin{equation}
\left\langle \theta^\mu\frac{\partial\mathcal{H}}{\partial\theta^\mu}\right\rangle = 
\frac{1}{Z}\int\dd^n\theta\int\dd^n p\:\exp(-\beta\mathcal{H}(\theta,p)) \theta^\mu\frac{\partial\mathcal{H}}{\partial\theta^\mu} = 
-\frac{1}{\beta Z}\int\dd^n\theta\int\dd^n p\:\theta^\mu\frac{\partial}{\partial\theta^\mu}\exp(-\beta\mathcal{H}(\theta,p)) = 
\frac{n}{\beta}
\end{equation}
after integration by parts and using that the trace $\partial\theta^\mu/\partial\theta^\mu = \delta^\mu_\mu = n$ gives the dimensionality $n$ of the parameter space. Analogously, 
\begin{equation}
\left\langle p_\mu\frac{\partial\mathcal{H}}{\partial p_\mu}\right\rangle = 
\frac{1}{Z}\int\dd^n\theta\int\dd^n p\:\exp(-\beta\mathcal{H}(\theta,p)) p_\mu\frac{\partial\mathcal{H}}{\partial p_\mu} = 
-\frac{1}{\beta Z}\int\dd^n\theta\int\dd^n p\:p_\mu\frac{\partial}{\partial p_\mu}\exp(-\beta\mathcal{H}(\theta,p)) = 
\frac{n}{\beta}
\end{equation}
again with the trace $\partial p_\mu/\partial p_\mu = \delta^\mu_\mu = n$. Both results apply regardless of the shape of the potential $\Phi$. We argue that the virial relation might serve as a convergence criterion for Markov-chains, with a well-defined value of $n$ for $\beta=1$ which is reached after equilibration, or burn-in. Naturally, derivatives of the Hamilton function $\mathcal{H}(\theta,p)$ with respect to the canonical momentum are trivial, with $T(p) = \delta^{\alpha\beta}p_\alpha p_\beta/2$ implying for the derivative $\partial T/\partial p_\mu = \delta^{\alpha\beta}(\delta^\mu_\alpha p_\beta + p_\alpha\delta^\mu_\beta)/2 = p^\mu$, such that the virial expression for the momenta becomes $\bra p_\mu p^\mu\ket = 2\bra T\ket$. It should be noted, however, that the validity of the virialisation condition does not require a kinetic energy which is quadratic in the momenta. The natural value of $\beta = 1$ for the inverse temperature for which the partition function falls back on the Bayesian evidence, then suggests that both virial expressions should become equal to the dimensionality.

\section{stationarity of the canonical ensemble}\label{sect:stationarity}
The thermal ensemble is stationary in the sense that the sampling from the posterior distribution $p(\theta,p|y)$ does not evolve with time. In fact, deriving cumulants $\kappa^m$ of the posterior $p(\theta|y)$ leads to
\begin{equation}
\kappa^{m} = 
\left.\frac{\partial^m}{\partial J^m}\ln Z[\beta,J_\alpha,K^\alpha]\right|_{K=0=J}
\quad\text{by differentiation of}\quad
Z[\beta,J_\alpha,K^\alpha] = \int\dd^n\theta\int\dd^np\:\exp(-\beta\mathcal{H}(\theta,p))\exp(\beta[J_\alpha\theta^\alpha + K^\alpha p_\alpha]).
\end{equation}
The time derivative of the cumulant is given by
\begin{equation}
\frac{\partial}{\partial t}\kappa^{m} =
\left.\frac{\partial^m}{\partial J^m}\frac{1}{Z}
\int\dd^n\theta\int\dd^np\:\exp(-\beta\mathcal{H}(\theta,p))\exp(\beta J_\alpha\theta^\alpha)\:J_\gamma\dot{\theta^\gamma}\right|_{J=0}
\end{equation}
as partial differentiations interchange. Here, we already discard the non-contributing terms involving $K^\alpha$. The time derivatives can be rewritten with the Hamilton equation of motion,
\begin{equation}
\dot{\theta}^\gamma = +\frac{\partial\mathcal{H}}{\partial p_\gamma}
\end{equation}
leading to
\begin{equation}
\frac{\partial}{\partial t}\kappa^{m} = -
\left.\frac{\partial^m}{\partial J^m}\frac{1}{\beta Z}
\int\dd^n\theta\int\dd^np\:\exp(\beta[J_\alpha\theta^\alpha])
\left[J_\gamma\frac{\partial}{\partial p_\gamma}\exp(-\beta\mathcal{H}(\theta,p))\right]\right|_{J=0}.
\end{equation}
Integration by parts then yields a vanishing integral,
\begin{equation}
\frac{\partial}{\partial t}\kappa^{m} = 
\left.\frac{\partial^m}{\partial J^m}\frac{1}{\beta Z}
\int\dd^n\theta\int\dd^np\:\exp(-\beta\mathcal{H}(\theta,p)) \: J_\gamma\frac{\partial}{\partial p_\gamma}\exp(\beta J_\alpha\theta^\alpha)\right|_{J=0} = 0,
\label{20}
\end{equation}
as $\exp(\beta{J_\alpha\theta^\alpha})$ does not depend on $p$, implying in summary that there is no time evolution of the configuration space cumulants. 

Likewise, the momentum space cumulants are given by
\begin{equation}
\kappa^{m}_T = 
\left.\frac{\partial^m}{\partial K^m}\ln Z[\beta,J_\alpha,K^\alpha]\right|_{K=0=J}.
\end{equation}
Their time derivative follows analogously
\begin{equation}
\frac{\partial}{\partial t}\kappa^{m}_T =
\left.\frac{\partial^m}{\partial K^m}\frac{1}{Z}
\int\dd^n\theta\int\dd^np\:\exp(-\beta\mathcal{H}(\theta,p))\exp(\beta K^\alpha p_\alpha)\:K^\gamma\dot{p_\gamma}\right|_{K=0}
\end{equation}
only using the other Hamilton equation of motion at this point
\begin{equation}
\dot{p}_\gamma = -\frac{\partial\mathcal{H}}{\partial \theta^\gamma}
\end{equation}
implying
\begin{equation}
\frac{\partial}{\partial t}\kappa^{m}_T = 
\left.\frac{\partial^m}{\partial K^m}\frac{1}{\beta Z}
\int\dd^n\theta\int\dd^np\:\exp(\beta[K^\alpha p_\alpha])
\left[K^\gamma\frac{\partial}{\partial \theta^\gamma}\exp(-\beta\mathcal{H}(\theta,p))\right]\right|_{K=0}.
\end{equation}
Again, integration by parts then yields a vanishing integral,
\begin{equation}
\frac{\partial}{\partial t}\kappa^{m}_T = 
-\left.\frac{\partial^m}{\partial K^m}\frac{1}{\beta Z}
\int\dd^n\theta\int\dd^np\:\exp(-\beta\mathcal{H}(\theta,p)) \: K^\gamma\frac{\partial}{\partial \theta^\gamma}\exp(\beta K^\alpha p_\alpha)\right|_{K=0} = 0,
\end{equation}
such that the cumulants becomes stationary. This result implies that after equilibration has set in, the Markov-chain samples from a stationary posterior distribution and that the cumulants do not evolve.

\section{equipartition and thermalisation}\label{sect:thermalisation}

\subsection{Eqipartition}
Equipartition is, in contrast to virialisation, characteristic of thermalised systems, whereas virialisation does not make assumptions about thermodynamic equilibrium. A straightforward calculation shows that the expectation values of the quantities $\theta^\mu\partial_\nu\Phi$
\begin{equation}
\left\langle \theta^\mu \frac{\partial\Phi}{\partial\theta^\nu}\right\rangle =
\frac{1}{Z}\int\dd^n\theta\int\dd^np\:\exp(-\beta \mathcal{H})\:\theta^\mu\frac{\partial\Phi}{\partial\theta^\nu} =
-\frac{1}{\beta Z}\int\dd^n\theta\int\dd^np\:\theta^\mu\frac{\partial}{\partial\theta^\nu}\exp(-\beta \mathcal{H}) =
\frac{1}{\beta Z}\int\dd^n\theta\int\dd^np\:\frac{\partial\theta^\mu}{\partial\theta^\nu}\:\exp(-\beta \mathcal{H}) = \frac{\delta^\mu_\nu}{\beta}
\end{equation}
and of $p_\mu\partial^\nu T$
\begin{equation}
\left\langle p_\mu \frac{\partial T}{\partial p_\nu}\right\rangle =
\frac{1}{Z}\int\dd^n\theta\int\dd^np\:\exp(-\beta \mathcal{H})\:p_\mu\frac{\partial T}{\partial p_\nu} =
-\frac{1}{\beta Z}\int\dd^n\theta\int\dd^np\:p_\mu\frac{\partial}{\partial p_\nu}\exp(-\beta \mathcal{H}) =
\frac{1}{\beta Z}\int\dd^n\theta\int\dd^np\:\frac{\partial p_\mu}{\partial p_\nu}\:\exp(-\beta \mathcal{H}) = \frac{\delta_\mu^\nu}{\beta}
\end{equation}
suggesting that the degrees of freedom become independent of each other. Furthermore, the expectation values are equal and proportional to temperature in equilibrium, from which we define a further convergence criterion for Markov-chains, for the specific value of $\beta = 1$.

It should be noted that equipartition is a much stronger condition than virialisation: Whereas virialisation sums over all degrees of freedom, equipartition makes a statement about the individual degrees of freedom of the system,
\begin{equation}
\left\langle\theta^\mu\frac{\partial\Phi}{\partial\theta^\mu}\right\rangle = 
\sum_{\mu\nu}\left\langle\theta^\mu\frac{\partial\Phi}{\partial\theta^\nu}\right\rangle = 
\sum_{\mu\nu}\frac{\delta^\mu_\nu}{\beta} =
\frac{n}{\beta}
\quad\text{and}\quad
\left\langle p_\mu \frac{\partial T}{\partial p_\mu}\right\rangle = 
\sum_{\mu\nu}\left\langle p_\mu \frac{\partial T}{\partial p_\nu}\right\rangle = 
\sum_{\mu\nu}\frac{\delta_\mu^\nu}{\beta} =
\frac{n}{\beta}.
\end{equation}
In consequence, virialisation is implied if equipartition is fulfilled. In addition, as the virialisation condition is built as an average over the equipartition conditions, fluctuations are suppressed according to the law of large numbers and the expectation value $n/\beta$ should be reached faster, again indicating that virialisation is the weaker criterion.

Additionally, the mixed expectation values turn out to be zero as coordinates and momenta are independent in Hamiltonian mechanics,
\begin{equation}
\left\langle \theta^\mu \frac{\partial T}{\partial p_\nu}\right\rangle =
\frac{1}{Z}\int\dd^n\theta\int\dd^np\:\exp(-\beta \mathcal{H})\:\theta^\mu\frac{\partial T}{\partial p_\nu} =
-\frac{1}{\beta Z}\int\dd^n\theta\int\dd^np\:\theta^\mu\frac{\partial}{\partial p_\nu}\exp(-\beta \mathcal{H}) =
\frac{1}{\beta Z}\int\dd^n\theta\int\dd^np\:\frac{\partial\theta^\mu}{\partial p_\nu}\:\exp(-\beta \mathcal{H}) = 0
\end{equation}
and similarly,
\begin{equation}
\left\langle p_\mu \frac{\partial\Phi}{\partial \theta^\nu}\right\rangle =
\frac{1}{Z}\int\dd^n\theta\int\dd^np\:\exp(-\beta \mathcal{H})\: p_\mu\frac{\partial\Phi}{\partial \theta^\nu} =
-\frac{1}{\beta Z}\int\dd^n\theta\int\dd^np\:p^\mu\frac{\partial}{\partial \theta^\nu}\exp(-\beta \mathcal{H}) =
\frac{1}{\beta Z}\int\dd^n\theta\int\dd^np\:\frac{\partial p_\mu}{\partial \theta^\nu}\:\exp(-\beta \mathcal{H}) = 0,
\end{equation}
illustrating the fact that the sampling in configuration space and momentum space is independent. Again, this characteristic of thermal equilibrium can be investigated in the burn-in of Markov-chains.

\subsection{Gelman-Rubin criterion as a particular case}
The Gelman-Rubin criterion \citep{gelman_rubin, brook_gelman, gelman_rubin2} quantifies convergence in Monte Carlo Markov-chains by comparing the (co)-variance generated by a single chain in its evolution with the (co)-variance of an ensemble of chains at the same instant. In ergodic cases, the two averages should coincide, and if properly equilibrated, the variance does not evolve anymore.

While the Gelman-Rubin criterion \citep[for reviews, see][]{2011, 2018arXiv181209384V} clearly quantifies stationarity, it is remarkable that a criterion based on (co)-variance alone is sufficient to ensure that the sampling is representative of the posterior distribution. The physical interpretation of the Gelman-Rubin criterion, however, seems to be identical to equipartition for Gaussian distributions: In fact, choosing a parabolic $\Phi(\theta) = \chi^2(y|\theta)/2$ typical for linear models
\begin{equation}
\Phi(\theta) = \frac{1}{2} F_{\alpha\beta}\theta^\alpha\theta^\beta
\quad\rightarrow\quad
\frac{\partial\Phi}{\partial\theta^\nu} = 
\frac{F_{\alpha\beta}}{2}\left(\delta^\alpha_\nu\theta^\beta + \theta^\alpha\delta^\beta_\nu\right) = 
F_{\alpha\nu}\theta^\alpha
\end{equation}
allows rewriting of the covariance (for simplicity, we have assumed centralised covariances) as
\begin{equation}
F_{\alpha\nu}\langle\theta^\mu\theta^\alpha\rangle = 
\left\langle\theta^\mu\frac{\partial\Phi}{\partial\theta^\nu}\right\rangle = 
\frac{\delta^\mu_\nu}{\beta}
\quad\rightarrow\quad
\langle\theta^\mu\theta^\nu\rangle = F^{\mu\nu}
\end{equation}
which comes out as the inverse Fisher-matrix at unit $\beta$ and is the basis of the well-know Fisher-matrix formalism in cosmology \citep{tegmark_karhunen-loeve_1997}. In consequence, monitoring the covariance in the spirit of the Gelman-Rubin criterion or the equipartition condition for the corresponding degree of freedom is equivalent. At this point it should be noted, that the Gelman-Rubin criterion compares two variances and is formulated as a statistical test for equality of two variances, i.e. the effectively it corresponds to a $t$-test, as both variances are statistically fluctuating quantities. In contrast, virialisation, equipartition and thermalisation make statements about an expectation value with a physically defined target value in thermal equilibrium. The equivalent formulation as a statistical test would take on the shape of an $F$-test for equal mean.

\subsection{Thermalisation and the exchange of energy with the canonical heat bath}
Driven by physical intuition one might keep a record of the thermal energy transferred to and dissipated from the system in the sampling process, where equilibration would be characterised by no net exchange of energy with the heat bath defining temperature: Initialising the Markov-chain close to the minimum position of the potential, which corresponds to the best fit point, would require an investment of energy for equilibration, and initialisation far away from the minimum position would necessitate that energy is dissipated until the equilibrium value of $n/\beta$ is reached. As the sampling is carried out at unit (inverse) temperature $\beta = 1$, the expectation value of the energy is given by $n$ directly. In equilibrium one would expect that the energy exchange with the heat bath fluctuates around an expectation value of zero.

It is important to notice that the exchange of thermal energy in burn-in takes place outside thermal equilibrium, such that drawing a connection to the change of thermodynamic entropy $\dd S$ as the reversibly exchanged heat normalised by the equilibrium temperature is difficult, simply because there is no notion of temperature outside equilibrium. This criterion is, in addition, attractive from a technical point of view: Keeping track of the energy exchange while sampling is a straightforward addition to a Markov-chain implementation, allowing for a convergence criterion without calculating the derivative of the potential and solving the equation of motion.

\section{numerical results}\label{sect:supernovae}
We investigate physically motivated convergence criteria for Markov-chains with a Hamilton Monte Carlo-algorithm, which samples efficiently microstates $(p_\mu,\theta^\nu)$ from the canonical partition sum
\begin{equation}
Z[\beta,J_\alpha,K^\alpha] = 
\int\dd^n\theta\:\int\dd^np\:\exp\left(-\beta\left[\frac{1}{2m}\delta^{\mu\nu}p_\mu p_\nu + \frac{\chi^2(y|\theta)}{2} + \phi(\theta)\right]\right)\exp\left(\beta\left[J_\alpha\theta^\alpha + K^\alpha p_\alpha\right]\right),
\end{equation}
i.e. the Hamiltonian function $\mathcal{H}(p,\theta) = T(p) + \Phi(\theta)$ separates into a conventional quadratic kinetic part and a potential,
\begin{equation}
T(p) = \frac{1}{2m}\delta^{\mu\nu}p_\mu p_\nu
\quad\text{as well as}\quad
\Phi(\theta) = \frac{\chi^2(y|\theta)}{2} + \phi(\theta).
\end{equation}
Expectation values of any phase space function $g(p,\theta)$ can be estimated from the samples $(p_\mu^{(i)},\theta^{\nu,(i)})$ provided by the Markov-chain
\begin{equation}
\langle g(p,\theta)\rangle = 
\int\dd^n\theta\int\dd^np\:p(\theta,p|y)g(p,\theta) = 
\frac{1}{Z}\int\dd^n\theta \int\dd^np\:\exp{(-\beta \mathcal{H}(p,\theta))}\:g(p,\theta) 
\approx 
\frac{1}{N} \sum_{i=1}^N g(p^{(i)},\theta^{(i)}).
\end{equation}
For instance, equipartition conditions in the previous section would be computed as 
\begin{equation}
\left\langle \theta^\mu \frac{\partial \mathcal{H}}{\partial \theta^\nu} \right\rangle 
\approx 
\frac{1}{N} \sum_{i=1}^N \theta^{\mu,(i)} \frac{\partial\Phi}{\partial \theta^{\nu}}(\theta^{(i)})
\label{eqn:approx_part}
\end{equation}
where the gradient $\partial\Phi/\partial\theta^\nu$ at the position $\theta^{(i)}$ can be evaluated by finite differencing. We work with an analytical expression of the gradients of $\Phi$ in the example Sect.~\ref{sect_numerics_toy} and use autodifferentiability of the physics-informed neural network implementation in Sect.~\ref{sect_numerics_supernova}.

\subsection{First numerical experiments}\label{sect_numerics_toy}
\begin{figure}
	\centering
	\includegraphics[width =0.45\textwidth]{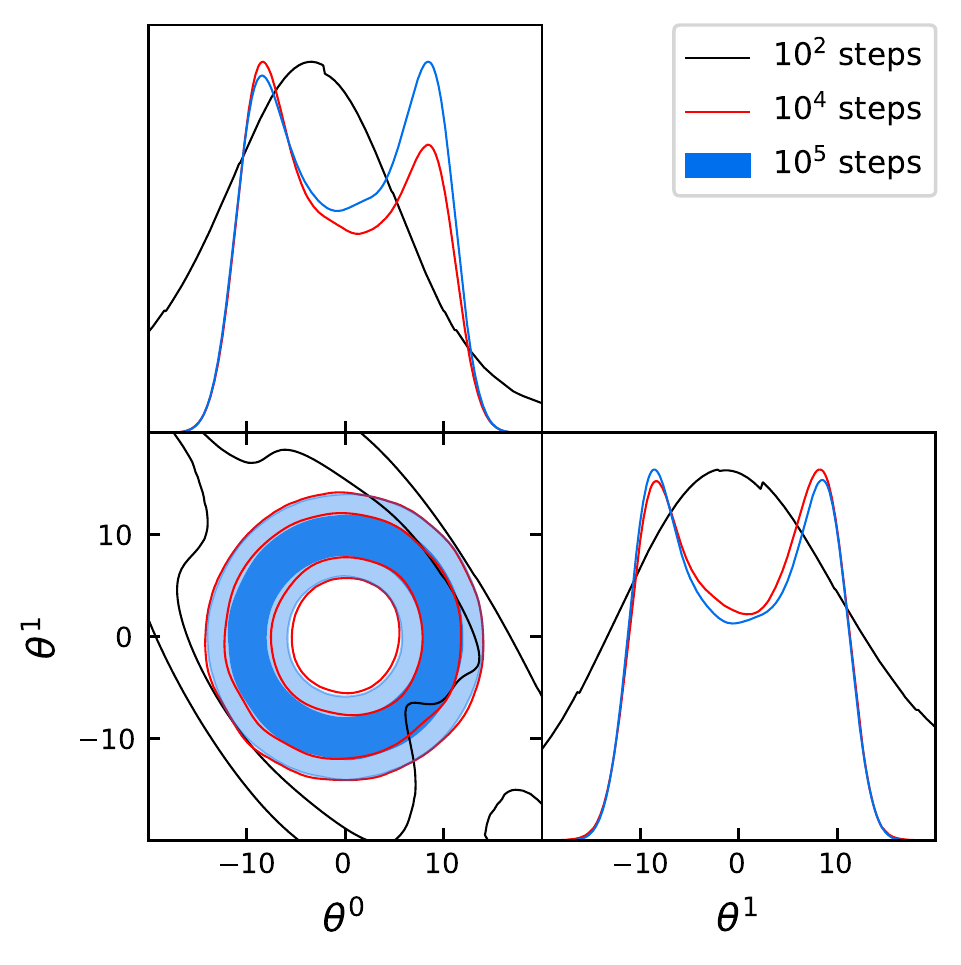}
	\includegraphics[width =0.45\textwidth]{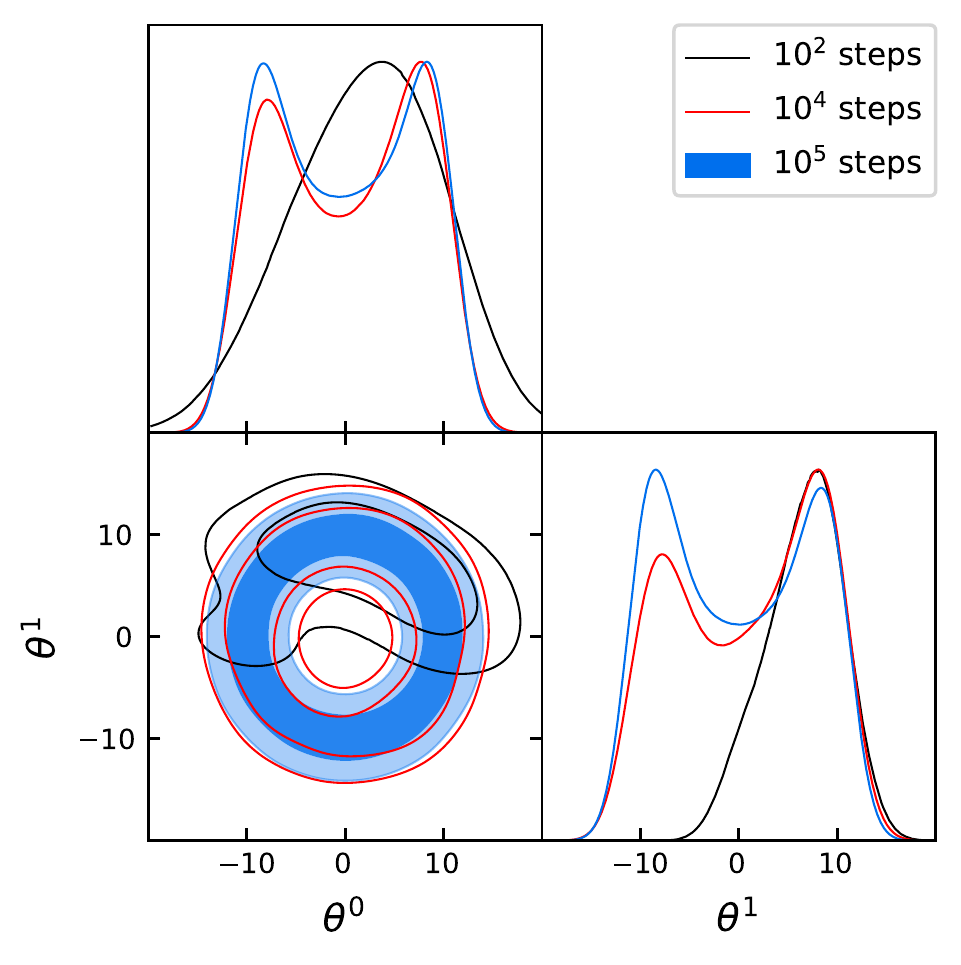}
	\caption{Kernel density estimates performed on the first $10^2$, $10^4$ and $10^5$ points of an HMC chain for the toy example. For the plot on the left initial conditions for the HMC where chosen away from the maximum posterior region, while for the plot on the right one of the most probable points was chosen as the initial condition.}\label{fig:ring_kdes}
\end{figure}

To demonstrate that the convergence criteria described in Sect.~\ref{sect:thermalisation} perform well in practice they are applied to a two-dimensional toy example with non-Gaussian shape and a strong degeneracy. The positions, associated momenta and derivatives of the potential are obtained using a basic Hamilton Monte Carlo- algorithm as described in \cite{Neal2011MCMCUH}. The likelihood sampled is of the form
\begin{equation}
	\mathcal{L} \left(\theta |R,\sigma\right) \propto 
	\exp\left(-\frac{\left(\sqrt{\theta_\nu\theta^\nu} - R\right)^2}{2\sigma^2}\right),
\quad\text{with the analytic derivative}\quad
	\frac{\partial}{\partial \theta^\mu} \left(- \ln \mathcal{L} \left(\theta |R,\sigma\right)\right) = \frac{\sqrt{\theta_\nu\theta^\nu}-R}{\sqrt{\theta_\rho\theta^\rho} \sigma^2}\theta_\mu.
\end{equation}
The Hamilton Monte Carlo-algorithm uses the derivatives of the potential to find the trajectories on which new points are proposed and estimates of the convergence criteria~(\ref{eqn:approx_part}) computed. Fig.~\ref{fig:ring_kdes} shows kernel density estimates, performed with getDist \citep{Lewis:2019xzd} on the  first $10^2$, $10^4$ and $10^5$ points of the Markov-chain, giving some intuition how well the chain reproduces the actual posterior after accumulating sufficiently many samples.

\begin{figure}
	\centering
	\includegraphics[width =0.45 \textwidth]{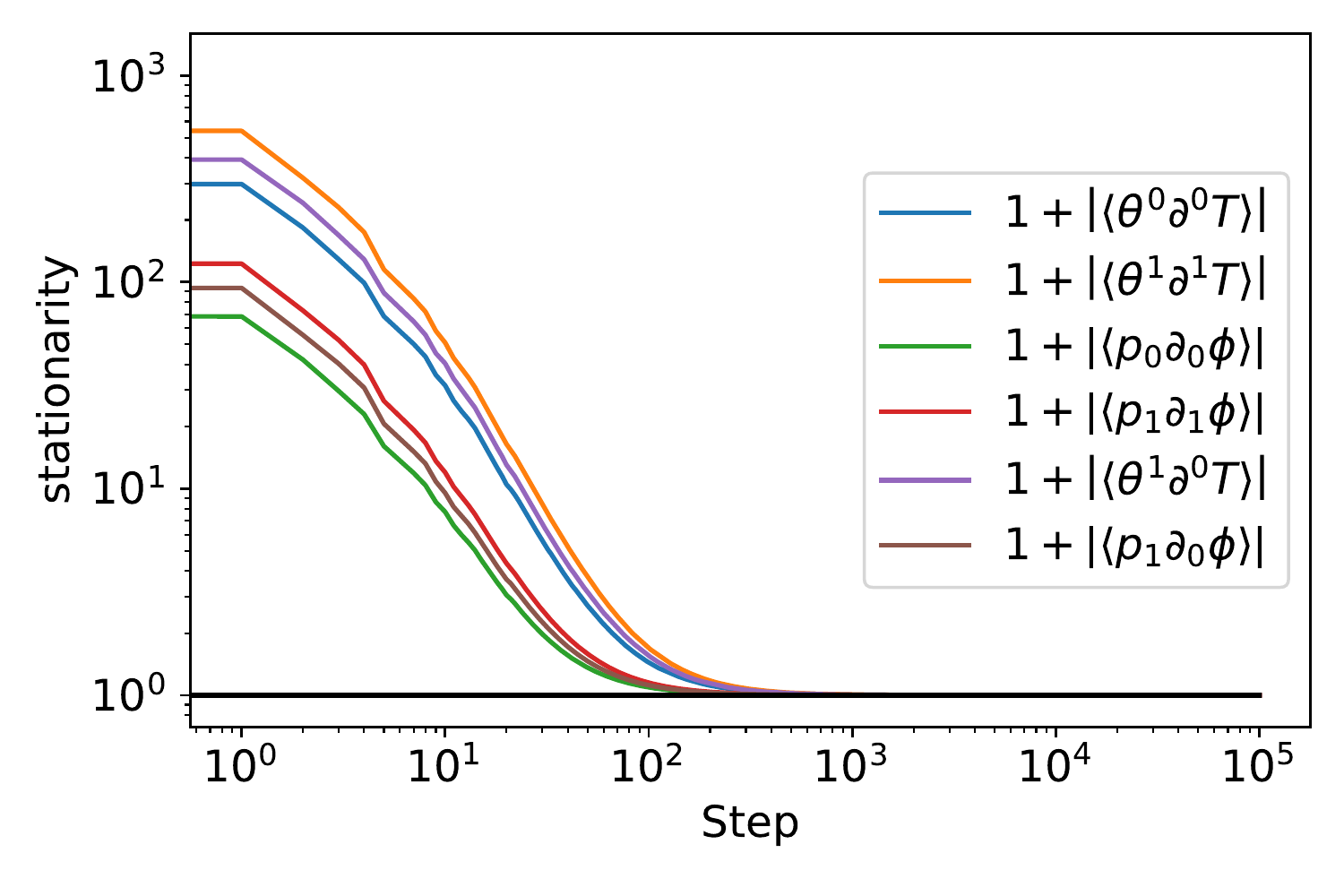}
	\includegraphics[width =0.45 \textwidth]{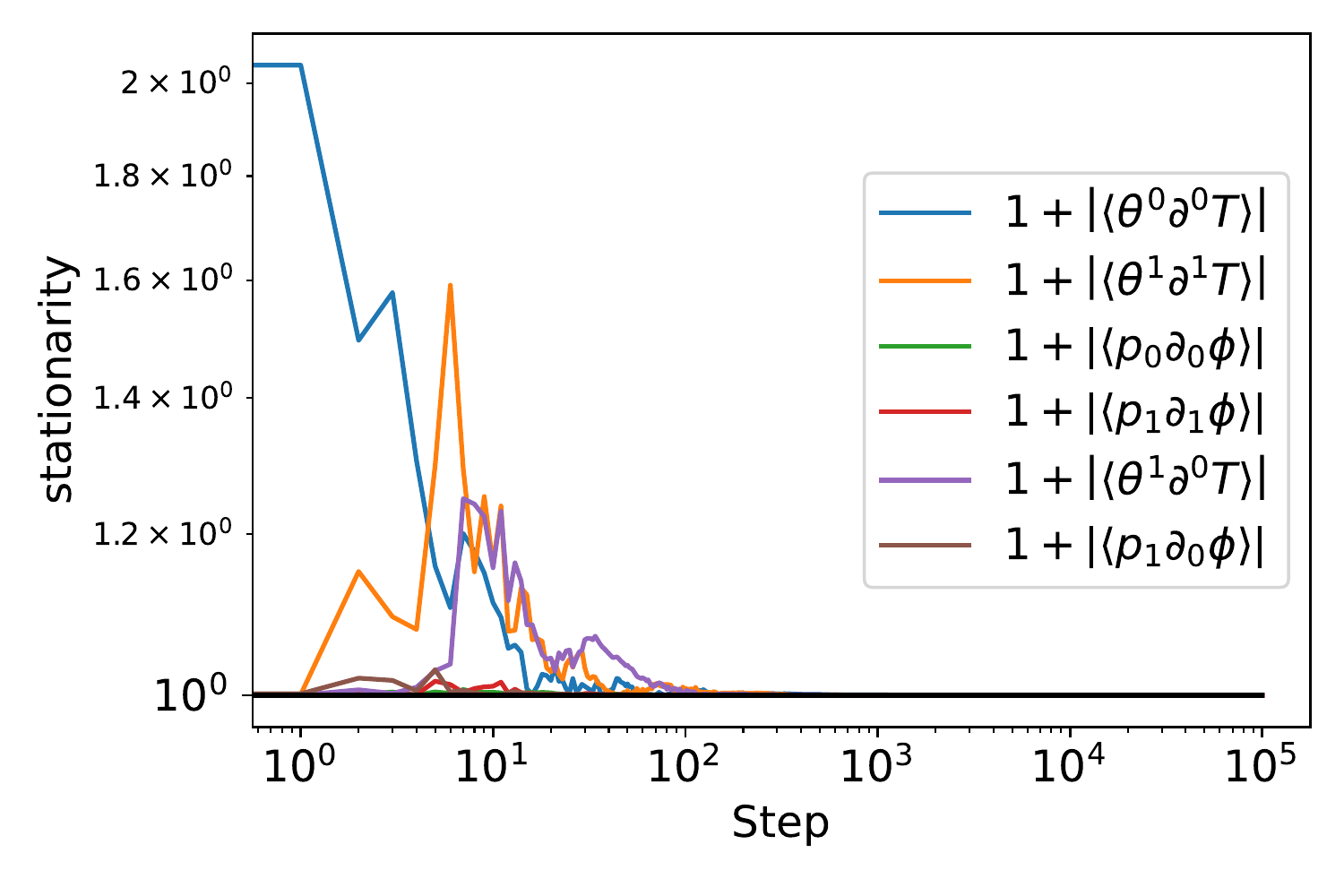}\\
	\includegraphics[width =0.45 \textwidth]{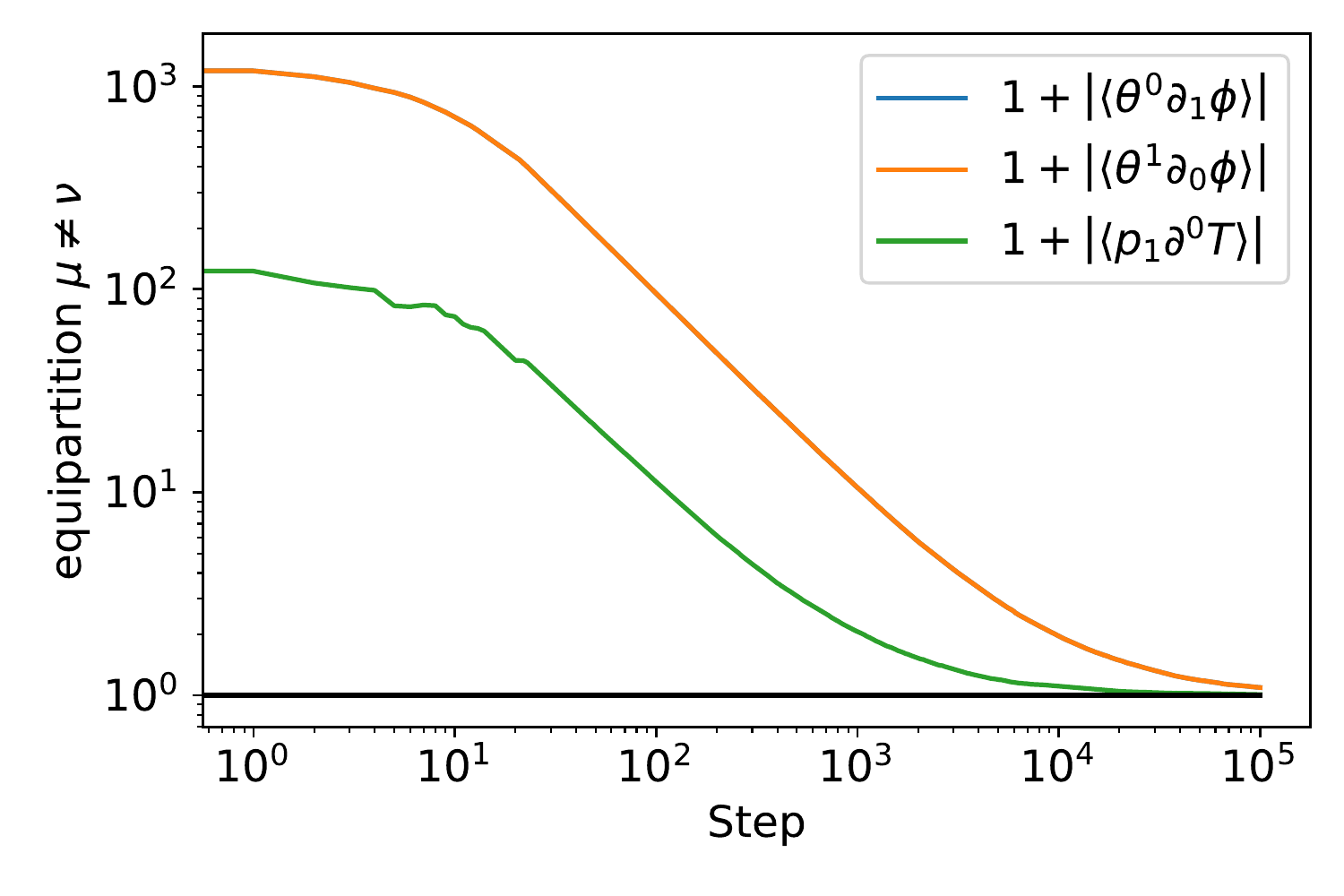}
	\includegraphics[width =0.45 \textwidth]{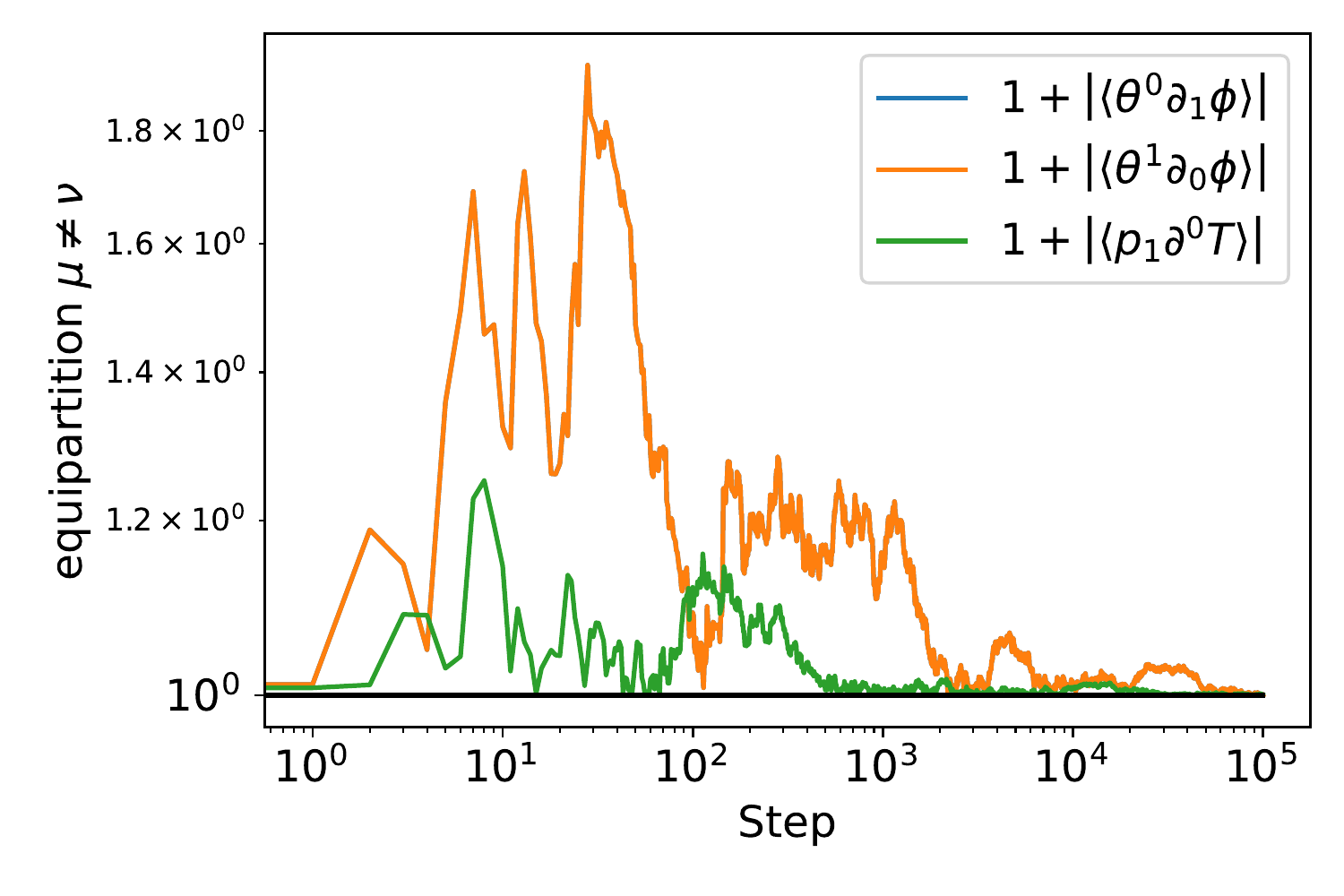}\\
	\includegraphics[width =0.45 \textwidth]{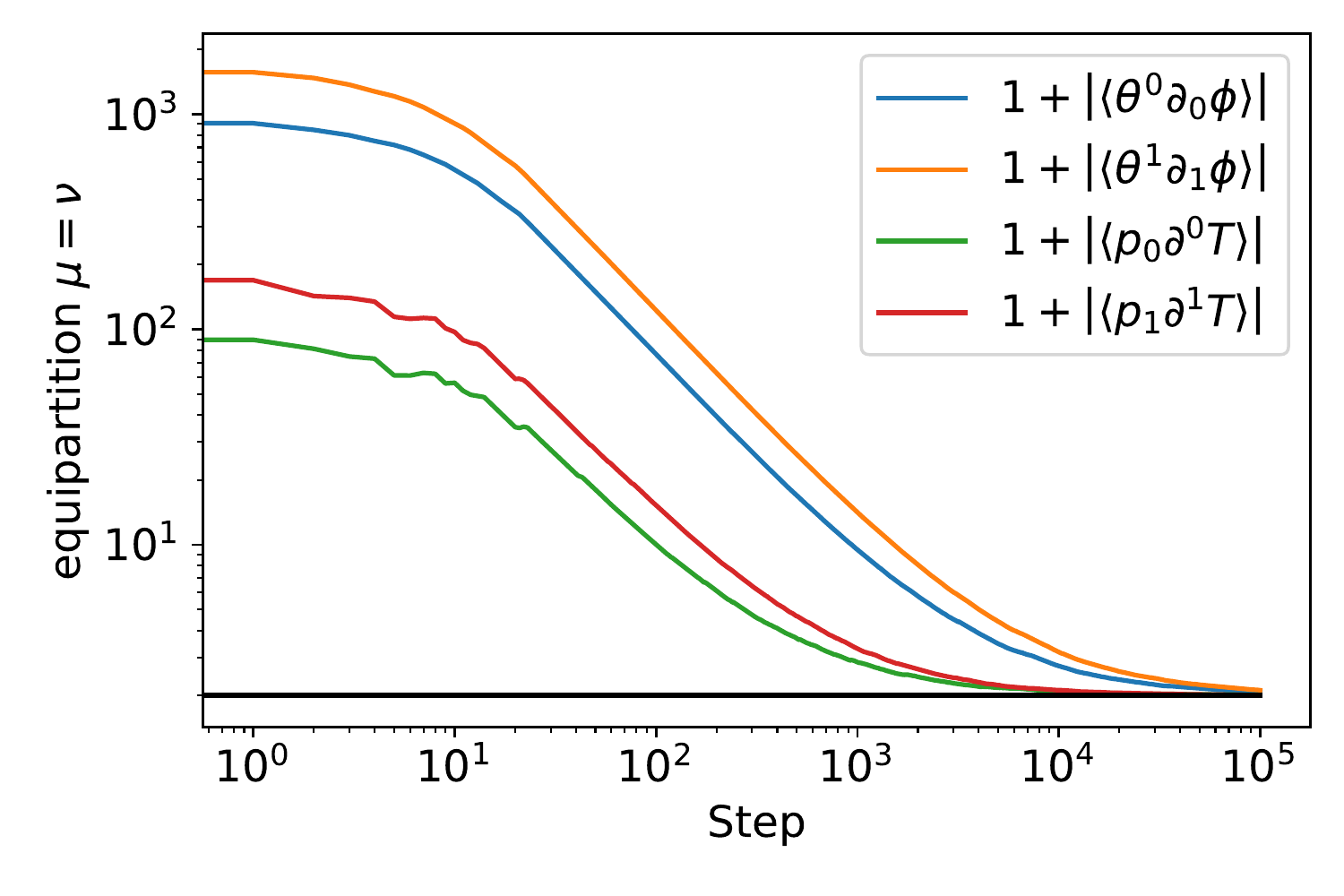}
	\includegraphics[width =0.45 \textwidth]{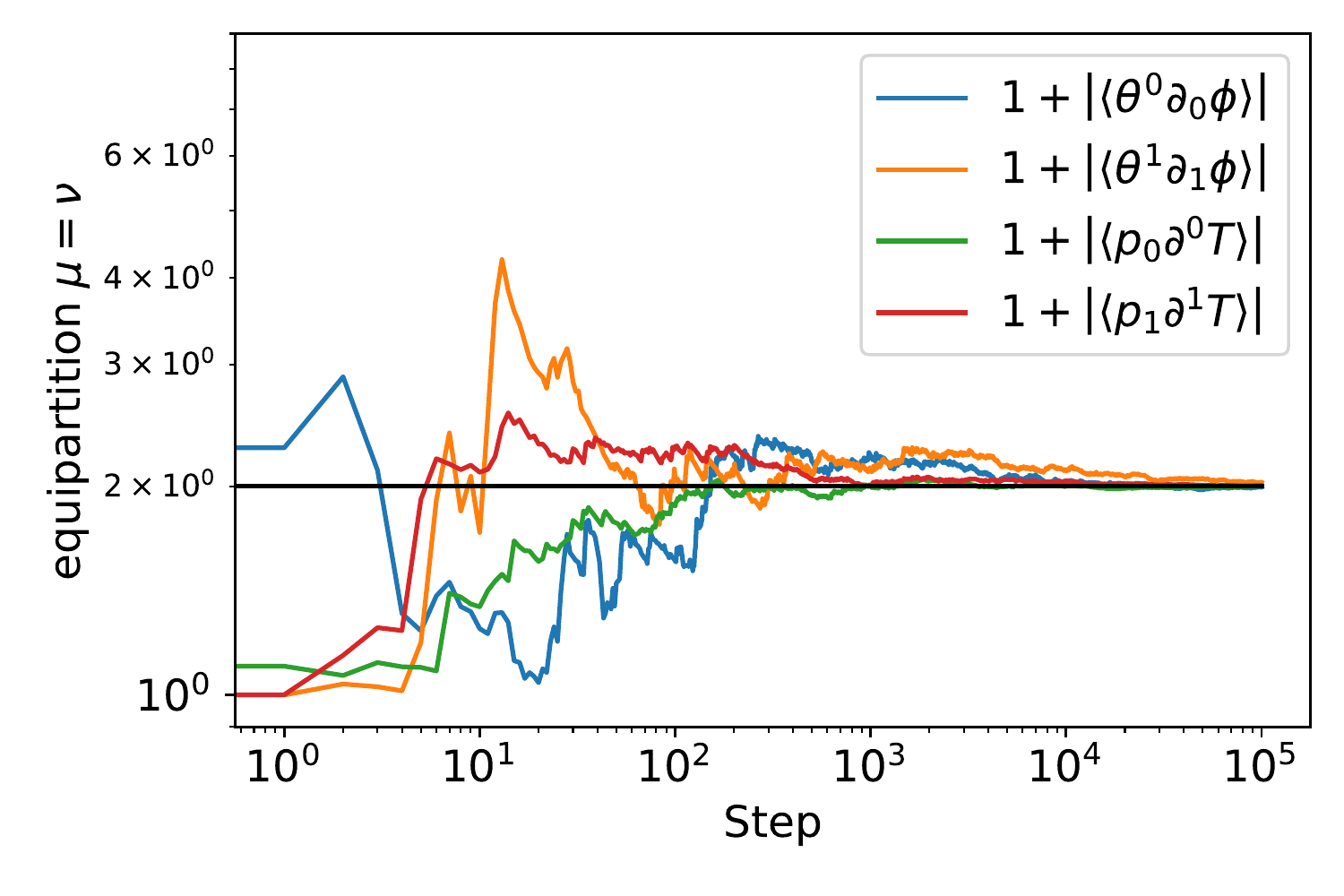}
	\caption{Progression plots of the stationarity condition and the equipartition of the different degrees of freedom in a HMC chain sampling the toy example. For the plots on the left initial conditions for the HMC-chain where chosen away from the maximum posterior region, while for the right column the maximum of the posterior was chosen as the initial condition.}
\label{fig:ring_convcrit}
\end{figure}

The cumulative values of the convergence criteria up to a specific step along the Markov-chain are shown in Fig.~\ref{fig:ring_convcrit}. For the left column of the figure the initial conditions of the Markov-chain were chosen far away from the minimum of the potential, whereas the initial conditions of the right column where at the (degenerate) minimum of the potential. The top row of plots illustrate the evolution of the mixed expectation value terms. When these go to zero in a system where approximate Hamiltonian dynamics are enforced, this signifies that the second moments of the probability distribution do not vary in time. Stationarity is realised surprisingly early in the evolution of the Markov-chain, even before $10^2$ steps are performed. In the centre row partition into different degrees of freedom is illustrated. The quantities $\langle\theta^\mu \partial_\nu \phi\rangle$ and $\langle p_\mu \partial^\nu T\rangle$, for $\mu \neq \nu$, tend towards zero as a larger amount of samples is accumulated. In these plots it is worth noting that the partition is significantly faster in the momentum degrees of freedom. This can be easily understood by recalling that the underlying distribution of the momenta is an uncorrelated normal distribution which is sampled from directly in the HMC algorithm. The lower row shows that the virial relations, i.e. the expectation values for $\mu = \nu$, tend towards one after a similar number of steps. While the left column illustrates the effect a long burn-in phase has on the different convergence criteria, the right column shows the effect of thermal fluctuations when the chain is started in a potential minimum. Even though we compute all expectation values cumulatively over all samples including those in the burn-in phase, a clear trend towards the thermal expectation values is seen which can help to quantify convergence.

\begin{figure}
	\centering
	\includegraphics[width =0.45\textwidth]{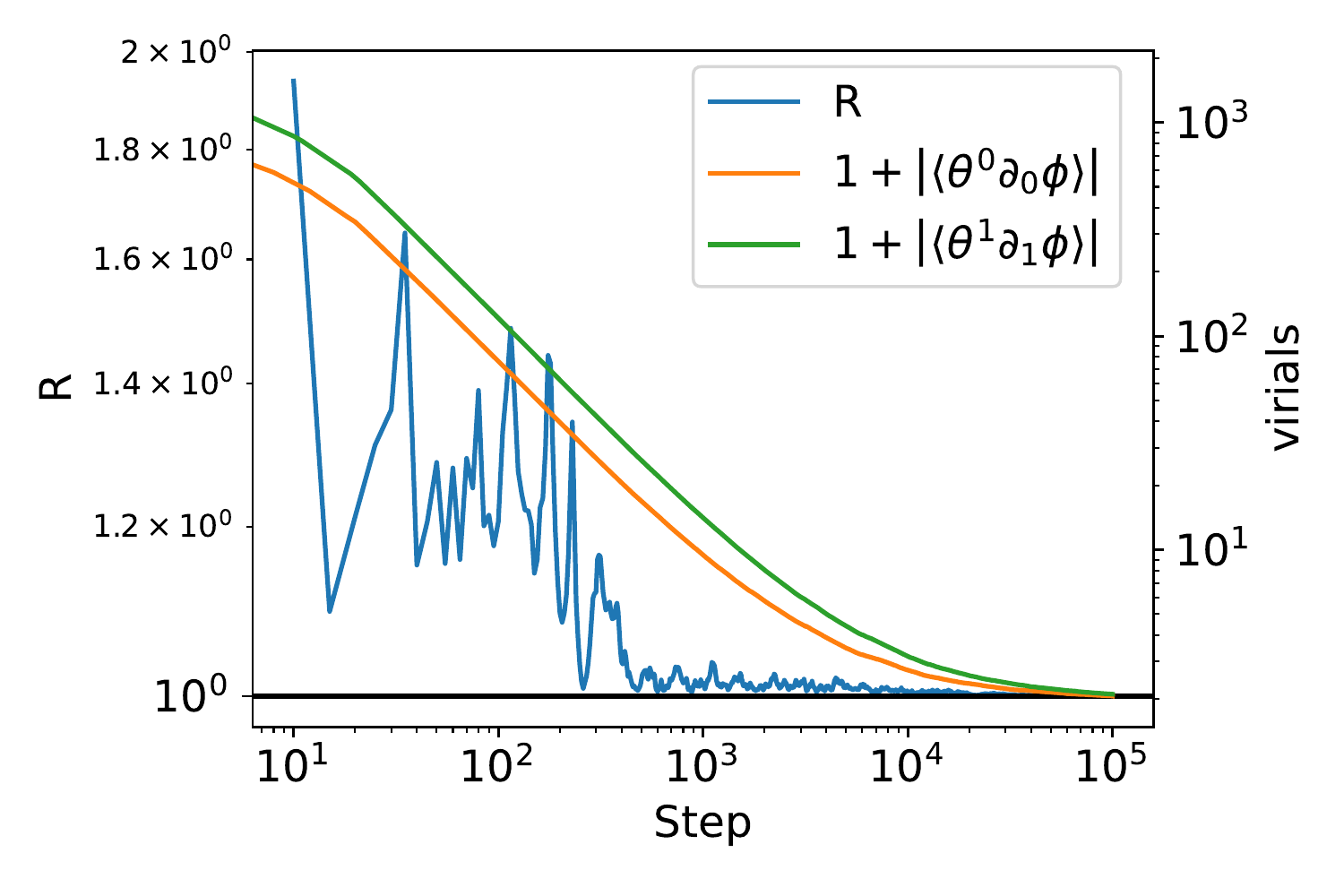}
	\includegraphics[width =0.45\textwidth]{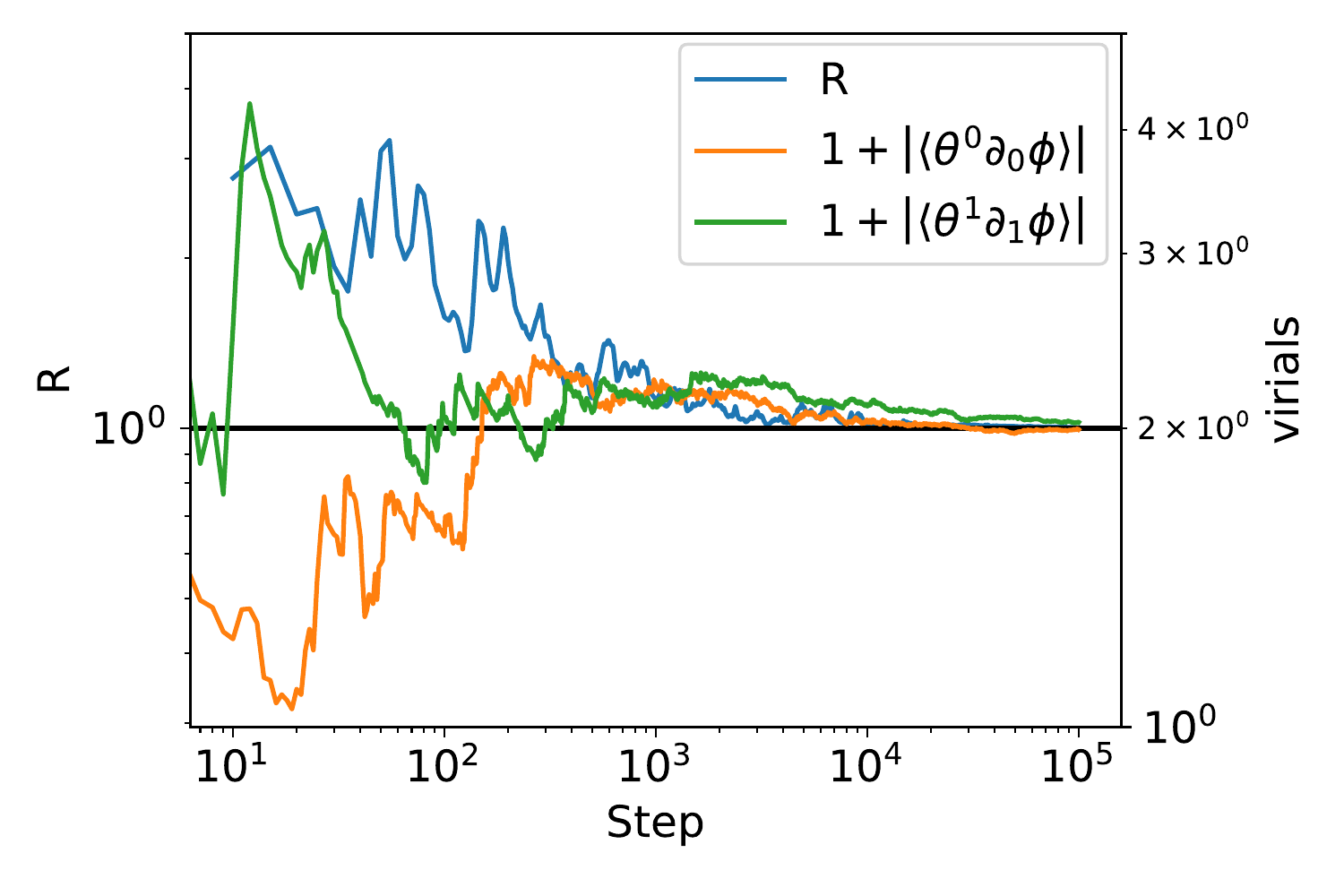}
	\caption{Comparison between the virialisation conditions and the Gelman-Rubin criterion $R$. For the ensemble averaging in the determination of the Gelman-Rubin criterion the Markov-chain was split into 10 batches. For the plot on the left initial conditions for the HMC where chosen away from the maximum posterior region, while for the plot on the right one of the most probable points was chosen as the initial condition.}\label{fig:ring_comparetoGR}
\end{figure}

Lastly, Fig.~\ref{fig:ring_comparetoGR} illustrates that the convergence of the Gelman-Rubin $R$ is commensurate with the virialisation conditions, in both cases of a well and badly chosen initial condition. Here we would like to emphasise that $R$ is a test statistic akin to a $t$-test and helps to decide between the hypothesis that the variances along a single Markov-chain and between an ensemble of independent Markov-chains are identical versus the hypothesis that this would not be true, at a selected confidence level. Similarly, one would quantify equality of the virialisation or equipartition conditions with the thermal expectation value by formulating a similar statistical test, in this case an $F$-test.

\subsection{Application to supernova data}\label{sect_numerics_supernova}
As a straightforward and relevant example for non-Gaussian likelihoods we consider constraints on the matter density $\Omega_m$ and the dark energy equation of state $\omega$ from the distance-redshift relation of supernovae of type Ia \citep{Riess1998, goobar_supernova_2011}. We impose a prior on spatial flatness and assume the equation of state to be constant in time. Constraints are derived from the Union2.1-data set \citep{suzuki_hubble_2012, amanullah_spectra_2010, kowalski_improved_2008}. The FLRW-distance modulus $y(z)$ as a function of redshift $z$ is given by
\begin{equation} \label{eqn:DistModulus}
y(z|\Omega_m,w) = 10 + 5 \log\left((1+z) \, \chi_H \int_{0}^{z}\dd{z^\prime}\:
\frac{1}{\sqrt{\Omega_m (1+z^\prime)^3 + (1-\Omega_m) (1+z^\prime)^{3(1+\omega)}}}\right).
\end{equation}
Constructing the likelihood for the two parameters $\Omega_m$ and $w$ for Gaussian errors $\sigma_i$ in the distance moduli $y_i$ yields a simplified expression, where we neglect correlations between the data points,
\begin{equation}
\likeli(y|\Omega_m,\omega) \propto \exp\left(-\frac{1}{2}\chi^2(y|\Omega_m,\omega)\right)
\quad\text{with}\quad
\chi^2(y|\Omega_m,\omega) = \sum_i\left(\frac{y_i - y(z_i|\Omega_m,\omega)}{\sigma_i}\right)^2.
\label{eqn:MCMC_likelihood}
\end{equation}
This likelihood is implemented in a Hamilton Monte Carlo sampler, with a uniform prior $\pi(\theta)$ for simplicity. For speeding up the computations, we employ physics-informed neural networks \citep[for an introduction, see][where details of our implementation are given in Appendix~\ref{appendix_pinn}]{DBLP:journals/corr/abs-1711-10561}. As the model prediction $y_i(z_i|\Omega_m,\omega)$ is given as an explicit function, we use an autodifferentiation functionality to derive gradients of $\chi^2(y|\Omega_m,\omega)$ needed in Hamilton Monte Carlo-sampling.

\begin{figure}
	\centering
	\includegraphics[width =0.45\textwidth]{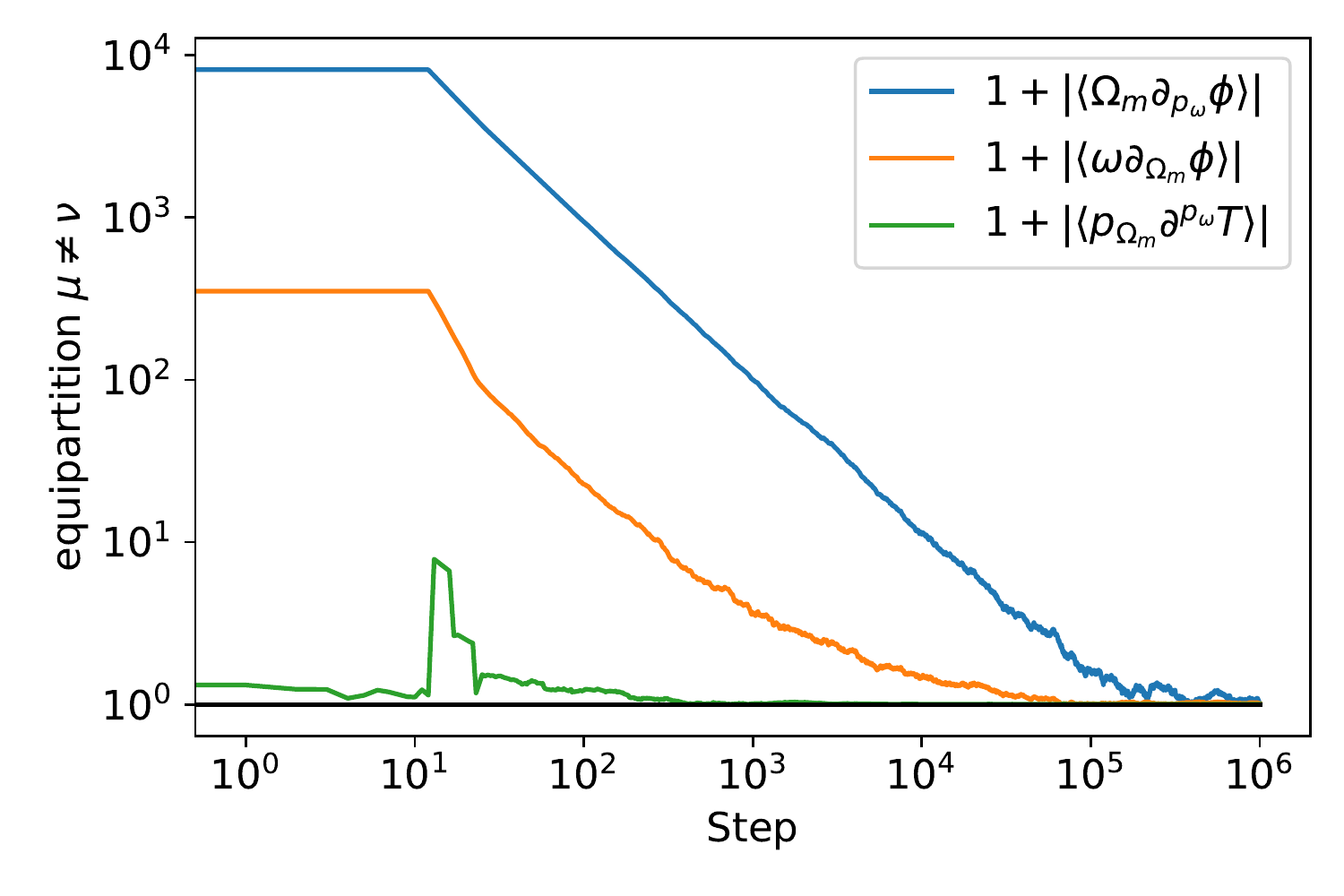}
	\includegraphics[width =0.45\textwidth]{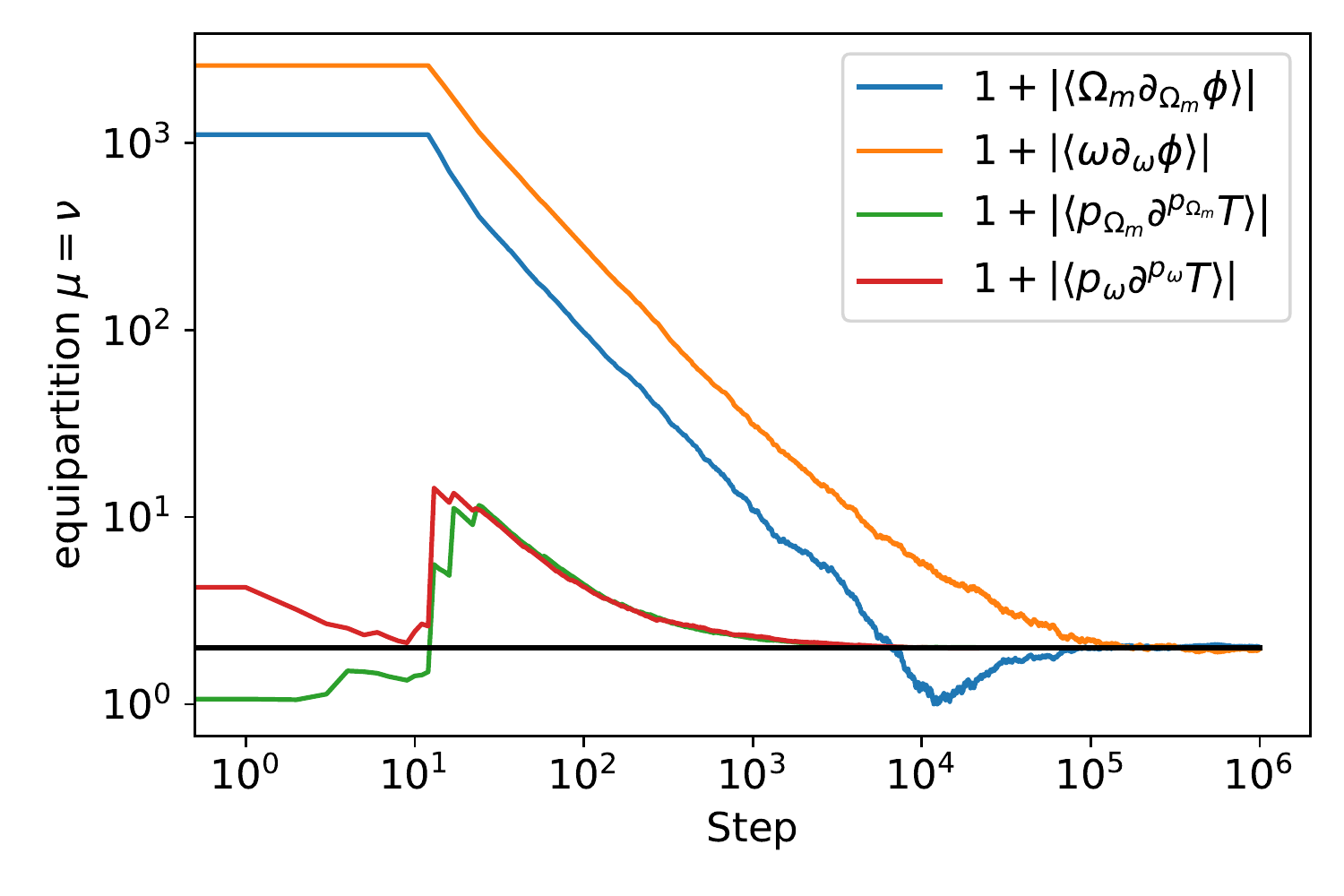}\\
	\caption{Application of the equipartition criterion to Hamilton Monte Carlo Markov-chains sampling the supernova likelihood. The left plot shows the partition into different degrees of freedom while the plot on the right shows that they are equipartitioned.}
	\label{fig:super_equipart}
\end{figure}

The convergence criteria discussed in the previous sections are applied to the PINN-enhanced supernova likelihoods in Fig.~\ref{fig:super_equipart}: There is a clear trend towards the values expected for thermal equilibrium, with a scaling $\propto \mathrm{step}^{-1}$ for the cumulatively computed values.

\begin{figure}
	\centering
	\includegraphics[width =0.45\textwidth]{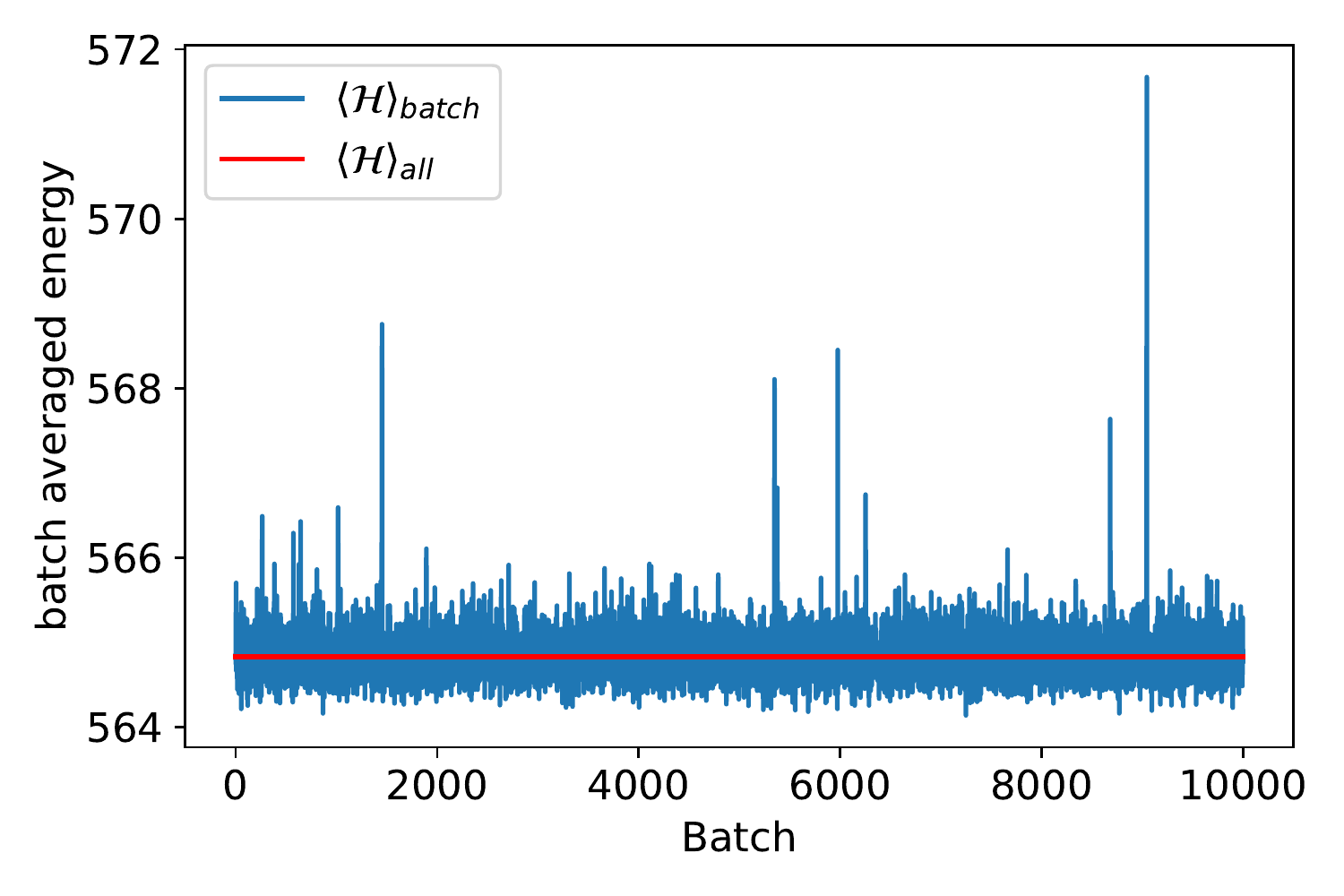}
	\includegraphics[width =0.45\textwidth]{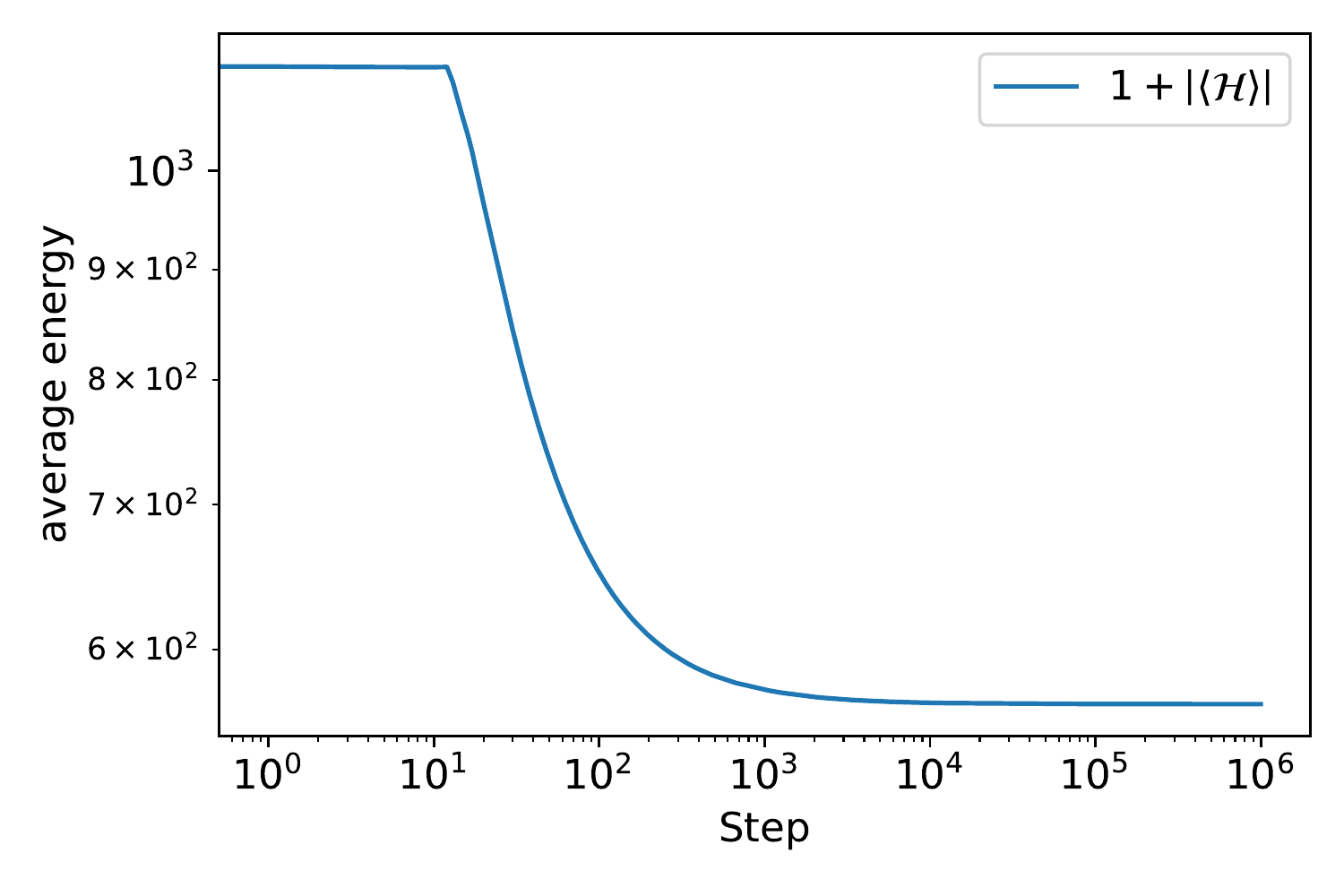}\\
	\caption{Application of the energy transfer convergence criteria to Hamilton Monte Carlo Markov-chains sampling the supernova likelihood. The plot on the left shows the averaged energy over 100 steps each, while the plot on the right shows the cumulative average at each step. }
	\label{fig:super_Etrans}
\end{figure}

Fig.~\ref{fig:super_Etrans} shows the average energy $\mathcal{H}(\theta,p)$ in the HMC system. For the left plot, energies are averaged over $10^2$ steps each and then, all $10^4$ batches are plotted successively. This allows to see the thermal fluctuations of these batch averages around an overall average value defined by the entire chain. The right plot depicts the cumulative average energy of the Markov-chain. The average energy in equilibrium can be estimated from its expected proportionality to the number of data points, as $y_i-y_\theta(z_i)$ in units of the error $\sigma_i$ is one on average, such that $\chi^2(y|\theta)$. Therefore, the average potential energy becomes equal to the number of data points. Clearly, one would need to take into account correlations between the data points, make sure that the distribution is Gaussian as well as rather consider the reduced $\chi^2$, which explains why the numerical value of $\langle\mathcal{H}\rangle$ falls short of the number of actual data points, which is 580.

\section{summary and discussion}\label{sect_summary}
The subject of this paper is the burn-in, equilibration phase of Monte Carlo Markov-chains and physical criteria by which it is possible to ascertain whether equilibration is reached. In equilibrium, Markov-chains provide samples out of a canonical ensemble constructed from the likelihood for a fit of a model to a data set together with the prior distribution, as an embodiment of Bayes' law. To this purpose, the Metropolis-Hastings algorithm and its variants set up a random process which consists of a thermal random walk in a potential derived from the logarithmic likelihood and the logarithmic prior. The burn-in phase of the Markov-chain consists of the first steps of this random walk, whose sample density, depending on initial position, is not a fair representation of the posterior distribution.

Commonly, convergence is characterised by the Gelman-Rubin criterion and the burn-in phase of the sampling discarded. The Gelman-Rubin criterion is a statistical test used to ascertain whether the inter-chain variance of a single Markov-chain is equal to the variance computed from an ensemble of chains, serving as a criterion of convergence. There is, as a hidden condition for the Gelman-Rubin criterion, the assumption of ergodicity, which is generally fulfilled in applications of the Metropolis-Hastings algorithm for continuous, bounded probability densities with respect to standard integral measures \citep{irreducible_markov, 10.1214/aos/1176325750}, in the sense that samples $\theta^{(i)}$ allow the approximate computation of any $p(\theta|y)$-weighted integral over a function $g(\theta)$ in the limit of many samples,
\begin{equation}
\lim_{N\rightarrow\infty}\frac{1}{N}\sum_{i=1}^N g(\theta^{(i)}) = \int\dd\theta\:p(\theta|y)g(\theta),
\end{equation}
i.e. the chain provides samples uniquely from the distribution $p(\theta|y)$. It is exactly this uniqueness that Gelman-Rubin quantifies, on the basis of the second moments of the sample distribution, for the choice $g(\theta) = \theta^2$. While certainly sensible, the Gelman-Rubin criterion quantifies only a single property of convergence, which motivated us to address the problem physically: The canonical partition sum $Z[\beta,J_\alpha,Z^\alpha]$ derived from the Bayesian evidence $p(y)$ represents the thermal motion of a particle in a potential at temperature $\beta$. Statistical physics provides straightforward quantitative predictions that characterise thermal equilibrium, which we investigate in Markov-chains converging towards their equilibria. In particular, we investigate convergence in Hamilton Monte Carlo chains because they resemble closed systems from statistical physics and have a notion of kinetic energy $T$ and potential energy $\Phi$, in addition to their many numerical advantages.

\begin{enumerate}
\item{All mechanical systems that are bounded in phase space fulfil a virialisation condition, where $\bra\theta^\mu\partial\Phi/\partial\theta^\mu\ket = \bra p_\mu\partial T/\partial p_\mu\ket$ applies on average. For the specific case of power-law potentials $\Phi\propto\theta^n$ and standard kinetic terms $T\propto p^2$ this implies $2\bra T\ket = n\bra\Phi\ket$ indicating the preference for a system for kinetic energy $n>2$ or for potential energy $n<2$. In thermal equilibrium, the expectation values can be computed to be $n/\beta$ with $n$ being the dimensionality of configuration space, which is a clearly defined criterion for $\bra\theta^\mu\partial\Phi/\partial\theta^\mu\ket$ and $\bra p_\mu\partial T/\partial p_\mu\ket$ to attain.}

\item{While the virialisation conditions are averaged statements, the equipartition conditions make a statement about individual degrees of freedom, specifically by demanding that $\bra\theta^\mu\partial\Phi/\partial\theta^\nu\ket\propto\delta^\mu_\nu$ and $\bra p_\mu\partial T/\partial p_\nu\ket\propto\delta^\nu_\mu$ with the proportionality constant being $1/\beta$, while $\bra p_\mu\partial\Phi/\partial\theta^\nu\ket = \bra p_\mu\partial T/\partial \theta^\nu\ket = 0$, thus defining the notion of a degree of freedom.}

\item{In thermal equilibrium there should be no energy exchange of a system with its environment: On average, the energy $\Delta\Phi$ provided by the heat bath at probability $\exp(-\beta\Delta\Phi)$ and the energy $-\Delta\Phi$ dissipated by the heat bath at unit probability should cancel each other.}

\item{As a physical application for Markov-chain convergence criteria we consider the posterior distribution $p(\Omega_m,\omega|y)$ derived from the magnitude-redshift relation of supernovae of type Ia. In order to speed up computations, we replace the integration of the luminosity distance by a physics-informed neural network, which is trained to yield the distance modulus $y(z|\Omega_m,\omega)$ for a given redshift $z$ for a wide range of possible cosmological models. Details of the physics-informed neural network implementation and its numerical accuracy are given in Appendix~\ref{appendix_pinn}.}
\end{enumerate}

In summary, virialisation conditions, equipartition conditions and the subsiding of the exchange of thermal energy are conceptually clear physical criteria with well-defined target quantities for thermal equilibrium, which could serve as measures of convergence for Markov-chains. Furthermore, they can be evaluated in ongoing  sampling processes with only a single Markov-chain. We have shown that the evolution of these quantities during burn-in shows similar properties as the Gelman-Rubin criterion. We would like to point out that there are in fact no-burn in methods for Markov-chains \citep{propp_wilson1996, Fill1997AnIA, fill2001extension}, which might yield advantages in time-consuming likelihood evaluations. We intend to continue these investigations for macrocanonical ensembles, where a constant particle number is an additional stationarity condition realised in equilibrium, as well as compare the performance of an anisotropic sampling process realising the partition function eqn.~(\ref{eqn_anisotropic_hmc}) to conventional Hamilton Monte Carlo samplers. In parallel, we intend to formulate statistical tests on the basis of the virialisation and equipartition conditions similar to the Gelman-Rubin statistic $R$ and investigate their equivalence.

\section*{acknowledgements}
This work was supported by the Deutsche Forschungsgemeinschaft (DFG, German Research Foundation) under Germany's Excellence Strategy EXC 2181/1 - 390900948 (the Heidelberg STRUCTURES Excellence Cluster).

\section*{data availability statement}
Our python-toolkit for monitoring virialisation, equipartition and thermalisation conditions of a Hamilton Monte Carlo sampler with physics-informed neural network speed-up (c.f. Appendix~\ref{appendix_pinn}) for the supernova likelihood is available on request.

\bibliographystyle{mnras}
\bibliography{references}

\appendix

\section{physics-informed neural networks and their application to supernova data}\label{appendix_pinn}
Physics-informed neural networks \citep[PINN,][]{DBLP:journals/corr/abs-1711-10561,DBLP:journals/corr/abs-2111-03794,DBLP:journals/corr/abs-2201-05624,hao2023physicsinformed} provide an approximation and interpolation to a parameterised set of functions, in our case the predictions for the distance modulus $y(z|\Omega_m,\omega)$ as functions of redshift $z$ parameterised by the dark matter density $\Omega_m$ and dark energy equation of state $\omega$. The model for the distance modulus described in eqn.~(\ref{eqn:DistModulus}) can be written in terms of the luminosity distance as 
\begin{equation}
	\mu = 5\log\left(d_L(z)\right) + 10.
\end{equation}
The integral expression for the luminosity distance can be expressed as an ordinary differential equation with given initial conditions
\begin{align}
	\frac{\dd d_L(z)}{\dd z}- \frac{d_L(z)}{1+z} - \frac{1+z}{H(z)} =0, \qquad d_L(0) = 0.
\end{align}
In this work we use a dense neural network of three hidden layers, each with a width of $50$ neurons. Similar to \cite{DBLP:journals/corr/abs-1711-10561} the loss function of the neural is composed of a term ensuring that the network output approximates the ordinary differential equation
\begin{equation}
	\mathcal{L}_\mathrm{ODE} = \frac{1}{N}\sum_{i=0}^{N}\left|\frac{\dd d_\mathrm{L,net}(z_i,\theta_i)}{\dd z}- \frac{d_\mathrm{L,net}(z_i,\theta_i)}{1+z_i} - \frac{1+z_i}{H(z_i,\theta_i)}\right|^2
\end{equation}
where the differentiation is performed using autograd. An additional term fixes the initial condition 
\begin{equation}
	\mathcal{L}_\mathrm{IC} = \frac{1}{N}\sum_{i=0}^{N}\left|d_\mathrm{L,net}(0,\theta_i)\right|^2.
\end{equation}
The overall loss function is given as the sum of the two components. The index $i$ denotes elements from the set $\{z_i, \theta_i\}_{i=0}^N$ of parameters generated uniformly over the relevant parameter ranges. The loss optimization was performed using PyTorch \citep{paszke2019pytorch} and the ADAM optimizer \citep{Kingma2014AdamAM} in a batch learning setup. 

\begin{figure}
	\centering
	\includegraphics[width = 0.45\textwidth]{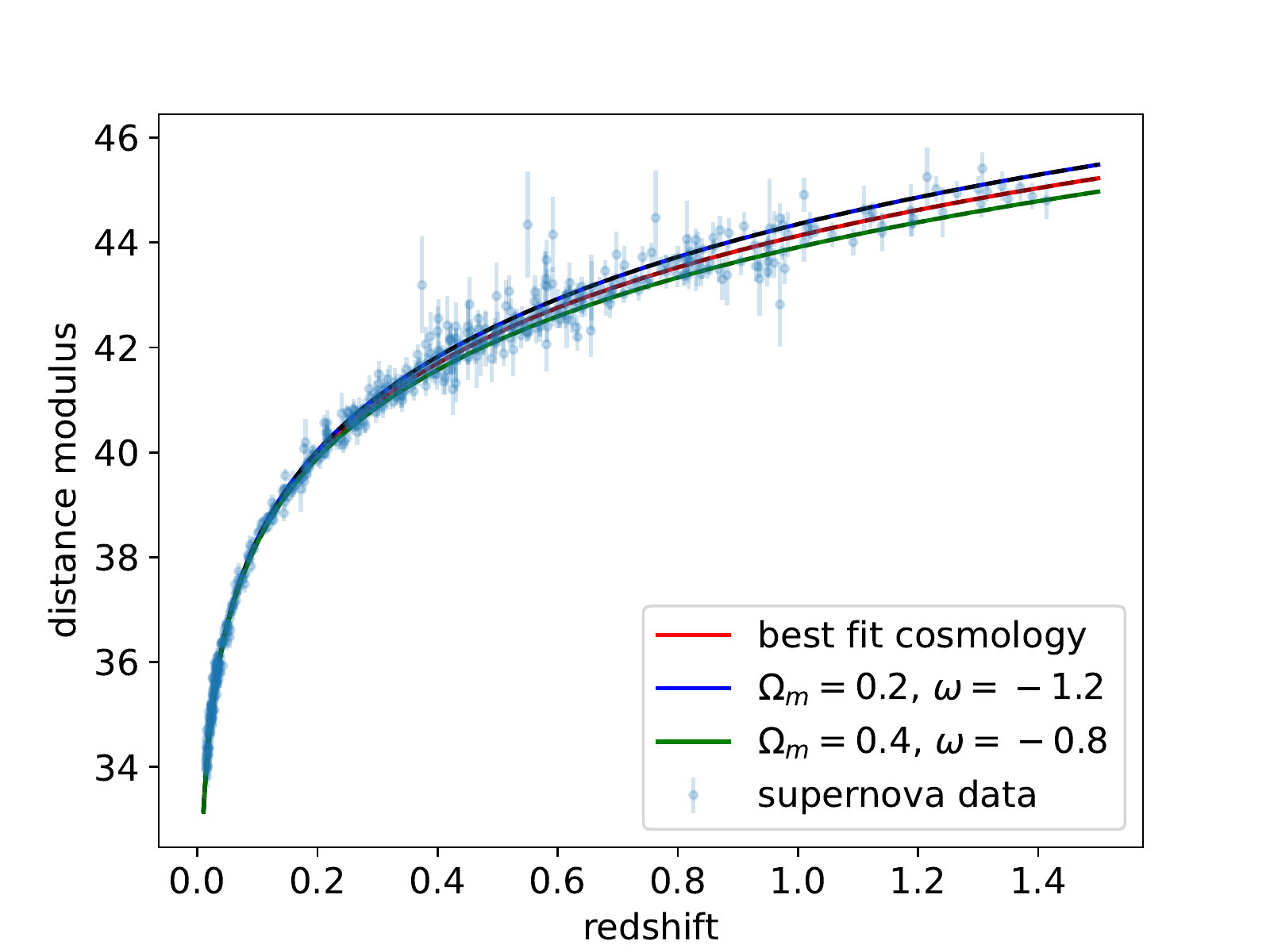}
	\caption{Comparison between distance moduli computed with the PINN and the differential equation formulation of (\ref{eqn:DistModulus}). Three different parameter combinations are shown, dashed lines are the PINN results.}\label{fig:PINNverify}
\end{figure}

Fig.~\ref{fig:PINNverify} shows the results from the trained PINN in comparison to the direct evaluation of $y(z|\Omega_m,\omega)$: By eye, the curves are virtually indistinguishable and the differences, amounting to less than a percent, are much smaller compared to the uncertainties in the supernovae's distance determination. Given these results, speeding up the $\chi^2$-evaluations in sampling by evaluating $y(z|\Omega_m,\omega)$ with the PINN seems justified. Furthermore, the gradients of $y(z|\Omega_m,\omega)$ with respect to the parameters which are necessary for the gradients of $\chi^2(y|\theta)$ in Hamilton Monte Carlo-sampling can be derived reliably, using automatic differentiation techniques.

\label{lastpage}
\end{document}